\documentclass[twocolumn]{aastex63}
\usepackage[normalem]{ulem}
\usepackage{amsmath}

\usepackage{lineno}
\newcommand{\tc}{t_c}

\newcommand{\gmax}{g_{\rm max}}

\newcommand{\Ms}{M_{\star}}

\newcommand{\OK}{\Omega_{\rm K}}

\renewcommand{\vec}[1]{\mathbf{#1}}
\newcommand{\pd}{\partial}
\newcommand{\pdt}[1]{\frac{\partial #1}{\partial t}}
\newcommand{\pdr}[1]{\frac{\partial #1}{\partial r}}
\newcommand{\pdp}[1]{\frac{\partial #1}{\partial \phi}}
\newcommand{\grad}[1]{\mathbf{\nabla} #1}
\newcommand{\avg}[1]{\left\langle #1 \right\rangle}

\newcommand{\be}{\begin{eqnarray}}
\newcommand{\ee}{\end{eqnarray}}

\newcommand\as{\bgroup\markoverwith{\textcolor[rgb]{.5, 0, .6}{\rule[0.5ex]{8pt}{1.5pt}}}\ULon}

\accepted{January 29, 2021}

\submitjournal{ApJ} 

\shorttitle{Rings formed by an eccentric instability}
\shortauthors{Li et al.}

\graphicspath{{./}{figures-production/}}

\begin{document}

\title{Ring Formation in Protoplanetary Disks Driven by an Eccentric Instability}

\correspondingauthor{Jiaru Li}
\email{jiaru\textunderscore li@astro.cornell.edu}

\author[0000-0001-5550-7421]{Jiaru Li}
\affiliation{Theoretical Division, Los Alamos National Laboratory, Los Alamos, NM 87545, USA}
\affiliation{Center for Astrophysics and Planetary Science,
Department of Astronomy, Cornell University, Ithaca, NY 14853, USA}

\author[0000-0001-8291-2625]{Adam M. Dempsey}
\affiliation{Theoretical Division, Los Alamos National Laboratory, Los Alamos, NM 87545, USA}

\author[0000-0003-3556-6568]{Hui Li}
\affiliation{Theoretical Division, Los Alamos National Laboratory, Los Alamos, NM 87545, USA}

\author[0000-0002-4142-3080]{Shengtai Li}
\affiliation{Theoretical Division, Los Alamos National Laboratory, Los Alamos, NM 87545, USA}

\begin{abstract}
We find that, under certain conditions, protoplanetary disks may spontaneously generate multiple, concentric gas rings without an embedded planet through an eccentric cooling instability.
Using both linear theory and non-linear hydrodynamics simulations, we show that a variety of background states may trap a slowly precessing, one-armed spiral mode that becomes unstable when a gravitationally-stable disk rapidly cools.
The angular momentum required to excite this spiral comes at the expense of non-uniform mass transport that generically results in multiple rings.
For example, one long-term hydrodynamics simulation exhibits four long-lived, axisymmetric gas rings.
We verify the instability evolution and ring formation mechanism from first principles with our linear theory, which shows remarkable agreement with the simulation results. 
Dust trapped in these rings may produce observable features consistent with observed disks. 
Additionally, direct detection of the eccentric gas motions may be possible when the instability saturates, and any residual eccentricity left over in the rings at later times may also provide direct observational evidence of this mechanism.
\end{abstract}


\section{Introduction}  
\label{sec:intro}

Observations of protoplanetary disks have revealed a diversity of substructures, ranging from annular rings \citep[e.g.,][]{ALMA2015,Andrews2016,Isella2016,Huang2018a}, to lopsided vortices \citep[e.g.,][]{Fukagawa2013,Marel2013} and spiral arms \citep[e.g.,][]{Benisty2015,Huang2018b}. 
Among these, the most common substructures are concentric rings and gaps \citep{Andrews2018,Garufi2018}, which may provide crucial insights to our understanding of protoplanetary disk evolution and planet formation.
Yet, their origins remain open questions. 

Unseen planets are commonly invoked to explain these features \citep[e.g.,][]{Dipierro2015,Dong2015,Dong2017,Jin2016,Zhang2018,WafflardFernandez2020}, as they may excite spiral waves \citep{Goldreich1979} and carve annular gaps in the disk under certain conditions \citep{Lin1986,Li2005}.
To produce rings at the observed location in their host disks, many proposed gap-opening planets must be at a few tens of AU (or even a hundred) from the star \citep[e.g.][]{Dipierro2015,Zhang2018}. 
However, without pre-exsiting substructures, dust and solid material can quickly drift from large radii to the inner parts of the disk \citep{Weidenschilling1977}. 
Given a low solid density environment and a long orbital timescale at large disk radii, the current core-accretion model of planet formation takes more than a million years to form planets \citep{Helled2014,Morbidelli2020}. 
Given this, some observed ringed disks may be too young to assemble the proposed planets \citep[e.g.,][]{ALMA2015,SeguraCox2020}. 

On the other hand, there are several ring-forming mechanisms that do not require planets.
These include a dust-gas viscous gravitational instability \citep{Tominaga2019}, 
and a magneto-hydrodynamical wind-driven instability \citep{Riols2019,Riols2020}.
Additionally, dust rings may occur at the edges of dead zones \citep{Flock2015} and snow lines \citep{Okuzumi2016}. 

Another interesting possibility is that rings may be formed via angular momentum exchange with an unstable non-axisymmetric mode. 
One candidate for this is an eccentric mode \citep{Lubow1991,Ogilvie2001,Lee2019a,Lee2019b}.
In this context, an eccentric mode refers to the coherent precession of gas on eccentric orbits across different radii.
The precession can be caused by both pressure and self-gravity.

It has been shown recently that almost any disk with a realistic density profile can sustain long-lived eccentric modes \citep{Lee2019b}. In non-self-gravitating disks, the linearized equation of motion for the complex eccentricity can be mapped to a Schr\"{o}nger-like equation, which shows that free eccentric eigenmodes can be trapped in the disk like a particle in a potential well \citep{Ogilvie2008,Saini2009,Lee2019b,Munoz2020}. These modes have discrete spectra and appear as apse-align elliptical rings. With disk self-gravity, \cite{Tremaine2001} and \cite{Lee2019a} found that disks can also support degenerate eccentric modes which appear as single-arm spirals. Forced eccentric modes have also been found to exist in disks coupled with both low-mass \citep[e.g.,][]{Papaloizou2002,Teyssandier2016} and high-mass secondary objects \citep{Munoz2020}.

A global eccentric mode can grow its amplitude through several mechanisms, such as: the SLING mechanism \citep{Adams1989,Shu1990} that amplifies an eccentric perturbation through the wobble of the central star, and instantaneous cooling \citep{Lin2015}.

In this paper, we present a theory of spontaneous (planet-free) gas ring formation through a spiral eccentric instability due to thermodynamic cooling. We find that this instability can produce a variety of gas features in just a few thousand orbits: from an initial spiral, to transient ellipses, and finally to a series of concentric rings. 

The rest of this paper is organized as follows. In Section~\ref{sec:Fid}, we present a hydrodynamical simulation that demonstrates the three substructures in one disk through its evolution. In Section~\ref{sec:Linear}, we formulate a linear theory for eccentric modes in disks with a finite cooling rate, and we use this theory to explain the growth and saturation of the spiral in the simulation. 
In particular, Section~\ref{sec:m0} discusses how an unstable one-armed spiral generically results in multiple, concentric rings. 
Section~\ref{sec:dis} discusses the astrophysical applicability and detectability of our mechanism. In Section~\ref{sec:Summary}, we give a summary of our findings.

\section{Fiducial evolution} \label{sec:Fid}

\subsection{Disk model}  \label{sec:Linear-disk}

We simulate a 2D self-gravitating gaseous disk with the hydrodynamics code \texttt{LA-COMPASS} \citep{Li2005,Li2008}. 
\texttt{LA-COMPASS} evolves the fluid equations with a Godunov, shock-capturing scheme.
In addition to evolving the continuity equation and momentum equations for the gas, \texttt{LA-COMPASS} also evolves the gas energy equation with an additional cooling term,
\begin{eqnarray}
\label{eq:Fid-adi}
    \frac{\partial E_g}{\partial t} + \grad \cdot\left( (E_g + P) \vec{v} \right) =  -\vec{v} \cdot \vec{\nabla} \Phi - \frac{c_v \Sigma (T - T_{\rm eq})}{t_c} ,
\end{eqnarray}
where $\Sigma$, $P$, $\vec{v}=u \hat{\vec{r}}+v \hat{\vec{\phi}}$, and $\Phi$ are the gas surface density, pressure, 2D velocity in the $r$-$\phi$ midplane, and total gravitational potential, respectively. 
The specific heat capacity is $c_v = k_b/(\mu (\gamma-1))$, $\gamma$ is the ratio of specific heats, $k_b$ is the Boltzmann constant, and $\mu$ is the mean particle mass which we fix to $\mu = 2.3 m_p$ where $m_p$ is the proton mass.
The total gas energy is given by $E_g = \Sigma |\vec{v}|^2/2 + P/(\gamma-1)$.
We assume that pressure is related to temperature, $T$, via the usual ideal gas equation of state, $P= (k_b/\mu) \Sigma T$. 

Gas cools to an equilibrium temperature $T_{\rm eq}$ on a timescale $t_c$ that is a fixed multiple of the orbital timescale given by the parameter $\beta = t_c \OK$, where $\OK$ is the Keplerian rotation rate \citep{Gammie2001}.
In the momentum equation, we include the gravitational potential of a central star with mass $\Ms$, as well as the disk's self gravity \citep{Li2009}. 
We do not include the indirect potential associated with the acceleration of the star due to the pull of an off-center disk. 
We have checked that a simulation including the indirect potential finds nearly exactly the same results as those presented in Section \ref{sec:Fid-picture}. 

The cooling model we adopt is meant to approximate the more accurate radiative cooling models thought to govern real protoplanetary disks.
We parameterize the cooling timescale with the relatively simple $\beta$ model that has been used extensively in studies of self-gravitating disks \citep[e.g.,][]{Gammie2001,Lodato2004,Lodato2005,Kratter2016}. 
This model has been shown to produce similar results to more sophisticated cooling models and makes an analytic treatment of our results possible in Section \ref{sec:Linear}.
Additionally, we also specify an equilibrium temperature profile rather than self-consistently computing one by balancing cooling with stellar and internal heating.
However, we do not expect this to impact our results significantly because more realistic equilibrium temperature profiles tend to be simple power-laws \citep[e.g.,][]{Chiang1997}.

While protoplanetary disks can be viscous in general, observations suggest many disks may have quite low viscosity \citep{Flaherty2017,Flaherty2018,Flaherty2020}. 
We focus on the inviscid limit by excluding an explicit disk viscosity term in the fluid equation, although there is implicit numerical viscosity in our simulations which is very small \citep{Li2009}.

\begin{figure}[b]
    \epsscale{1.1}
    \plotone{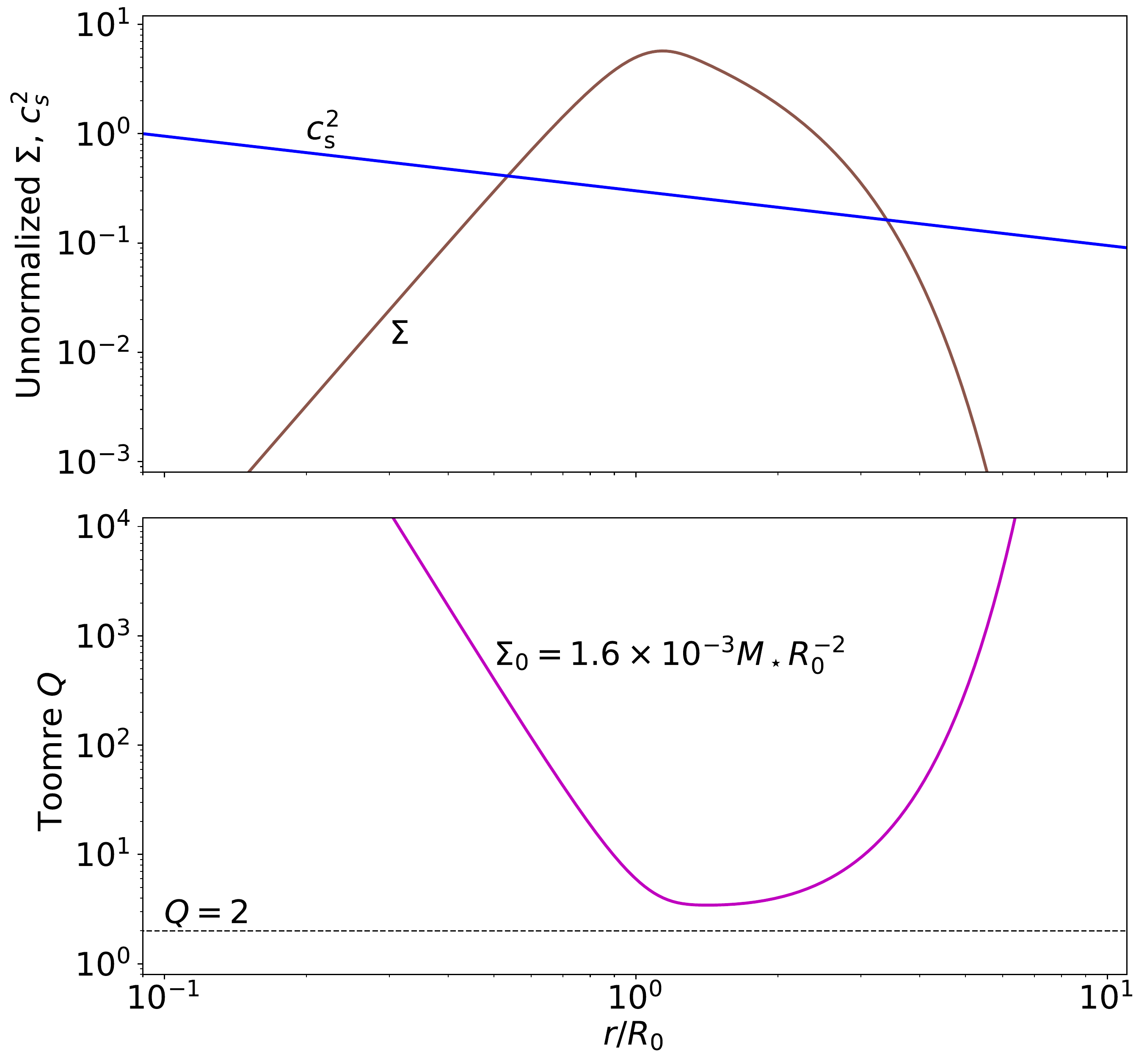}
    \caption{Initial $\Sigma$ and $c_{\rm s}^2$ (top) and radial $Q$ profiles (bottom) used in our study. Note that we only consider gravitationally stable disks with $Q > 2$.
    }
    \label{fig:Fid-profile}
\end{figure}

\subsection{Initial conditions}

For the initial surface density, we choose a power-law profile with both an inner hole and an outer taper,
\be
 \Sigma(r) = 2.03 \Sigma_0 \underbrace{\left(1 - e^{-(r/R_0)^6} \right)}_{\rm inner \, hole} \overbrace{\left(\frac{R_0}{r}\right)}^{\rm power-law} \underbrace{e^{-(r/(2 R_0))^2}}_{\rm outer \, taper} , 
\ee
We choose our reference radius to be $R_0$ and reference density to be $\Sigma_0 = \Sigma(R_0)$.

The initial temperature is set via the gas sound speed, $c_s^2(r) = \gamma (k_b/\mu) T = c_0^2 (r/R_0)^{-1/2}$.
We set $T_{\rm eq}$ to this initial, power-law temperature.
For all of our simulations we set $\gamma = 1.5$ and $c_0 = 0.03\sqrt{G M_\star/R_0}$.
Additionally, we set our unit of time to be the Keplerian frequency at $R_0$, i.e, $\OK(R_0)= \sqrt{G M_\star/R_0^3}$.

The simulations are performed in a two-dimensional cylindrical domain from $R_{\rm in}=0.2$ to $R_{\rm out}=6.0$ on a uniformly spaced $(N_{\rm r}, N_{\rm \phi}) = (4096, 4096)$ grid. The fluid field at the inner and outer boundaries is fixed at its initial value. 
A damping method for $r\in[0.2,0.3] \cup [5.6,6.0]$ is applied to prevent wave reflection \citep{volBorro2006}. We set the initial azimuthal velocity $v$ to provide the exact radial force balance, while a non-zero initial radial velocity $u$ is sampled uniformly from $[-10^{-3}, +10^{-3}]r\OK$ in each fluid cell as an initial perturbation. 
This perturbation could be caused by various processes, such as random fluid turbulence \citep{Balbus1998,Flaherty2020}.
We have checked that the disk evolution is independent of the type and amplitude of the initial perturbation so long as there is power in the $m=1$ azimuthal component.

\subsection{Fiducial evolution}
\label{sec:Fid-picture}

Our \textsc{Fiducial} disk adopts $\Sigma_0=1.6\times10^{-3} M_\star R_0^{-2}$ and $\beta=10^{-6}$. Hence, the disk has a total mass $0.0186\Ms$ and is essentially locally isothermal\footnote{Despite the name, essentially locally isothermal is fundamentally different from strictly locally isothermal, which does not evolve an energy equation, as the latter does not conserve wave angular momentum \citep{Lee2016,Miranda2019}.}. 
With radiative cooling, typical protoplanetary disks can have $\beta \ll 1$ \citep[][see also Section~\ref{sec:dis-beta}]{Zhang2020}.
The effect of $\Sigma_0$ and $\beta$ are explored in Section~\ref{sec:Linear-analysis}. 

Figure~\ref{fig:Fid-profile} shows our initial disk configuration.
The surface density peaks just outside of $R_0$ coinciding with a minimum in the Toomre $Q = c_s \Omega/(\pi G \Sigma)$ \citep{Toomre1964}. We focus in this paper on disks that are gravitationally stable, $Q > 2$, to ensure that any instability we find is not due to a small $Q$. 

\begin{figure}[t]
    \epsscale{1.1}
    \plotone{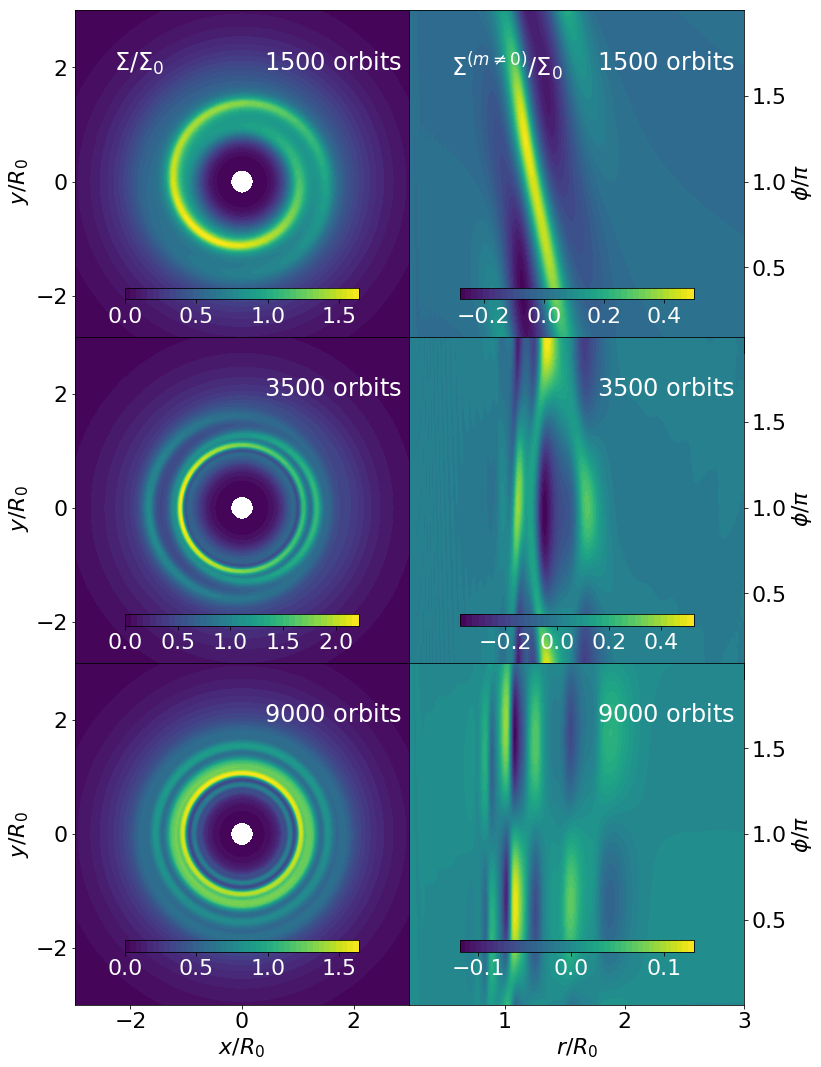}
    \caption{Snapshots of the \textsc{Fiducial} hydrodynamics simulation at time $=1500$ orbits (top), $3500$ orbits (middle), and $9000$ orbits (bottom). The left column shows disk surface density. The right column shows the $m\neq0$ component of the density deviation in the $r$-$\phi$ coordinates.
    }
    \label{fig:Fid-hydro-snapshot}
\end{figure}

\subsubsection{Modal time evolution}

We let the disk evolve for $10^{4}$ orbits at $R_0$ to provide a general picture of the long-term evolution. This corresponds to roughly
one million years around a solar mass star if $R_0 = 21.5$ AU.

Figure~\ref{fig:Fid-hydro-snapshot} shows three surface density snapshots. The disk evolves through three different stages: single-arm spiral (time $=1500$ orbits), multiple closed eccentric rings ($3500$ orbits), and multiple nearly circular rings ($9000$ orbits). These features are all radially confined between $r \sim 1-2 R_0$.

\begin{figure*}[htb!]
    \epsscale{1.1}
    \plotone{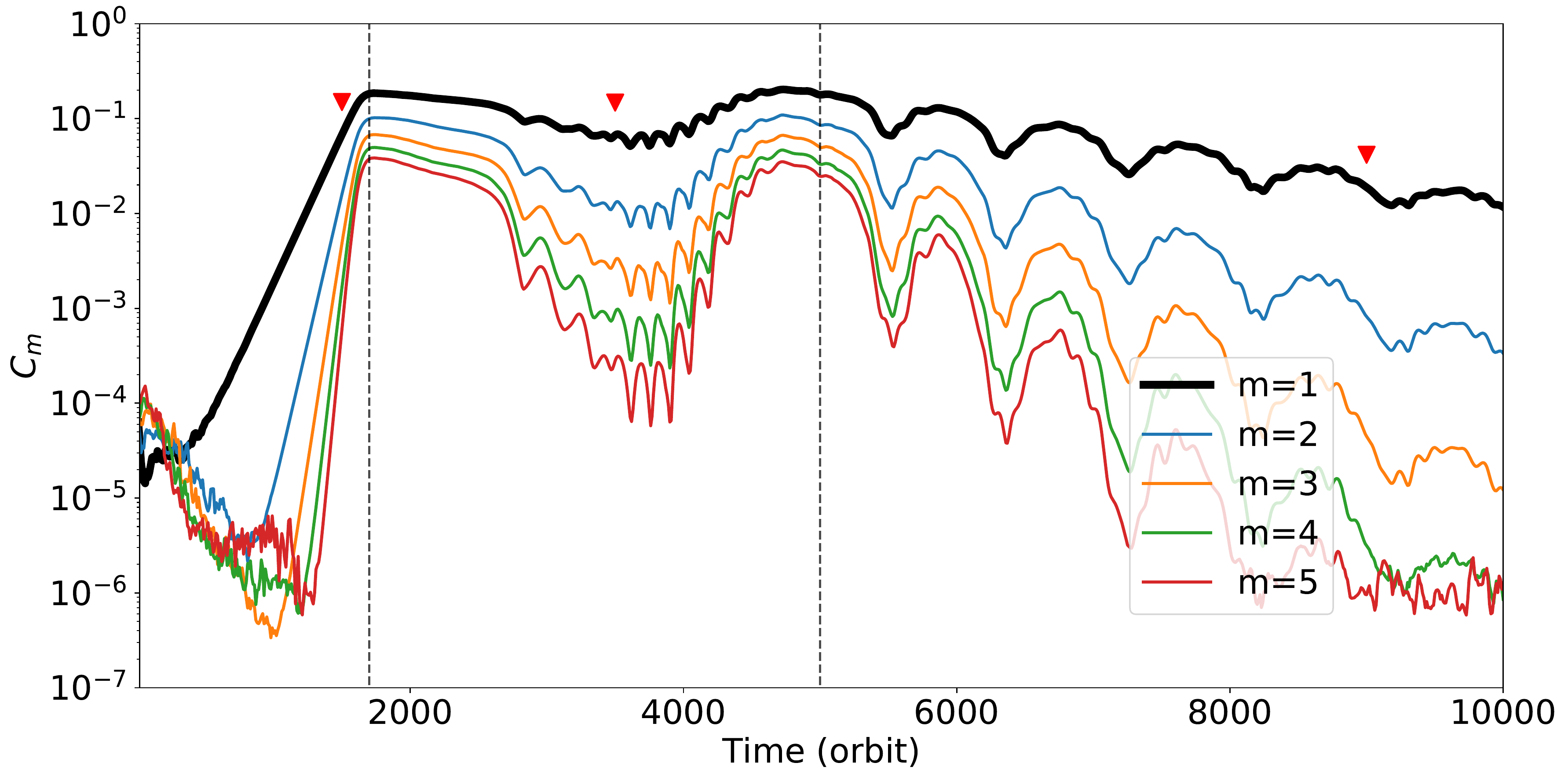}
    \plotone{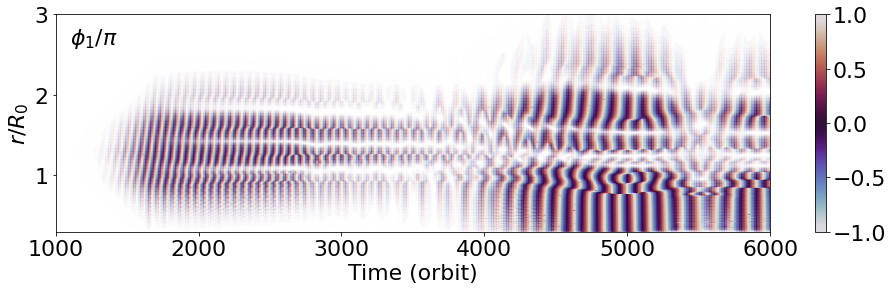}
    \caption{{\bf Top:} Time evolution of the global Fourier mode amplitudes for $m=1-5$ in the \textsc{Fiducial} simulation. The red triangles marks the time of the snapshots in Figure~\ref{fig:Fid-hydro-snapshot}. The dashed black vertical lines divide the whole evolution into (a) the initial growth of the spiral, (b) the saturation and morphology transition to closed eccentric rings, and (c) final multi-ring stage.
    {\bf Bottom:} Space-time diagrams of the local $m=1$ Fourier mode phase, $\phi_1$, from 1000 to 6000 orbits. The opaqueness of the colormap scales with the eccentricity at $(r,t)$. The $m=1$ pattern is precessing at a coherent rate of $\Omega_{\rm p}=0.018$ when its is a spiral and a different rate of $\Omega_{\rm p}=0.010$ after saturation.
    }
    \label{fig:Fid-hydro-Cm}
\end{figure*}

To get a sense of the relative importance of the different azimuthal symmetries in the disk, we show the time evolution of the global Fourier amplitudes,
\begin{eqnarray}
    C_m(t) \equiv \frac{1}{M_{\rm d}}\int_{R_{\rm in}}^{R_{\rm out}} \left|  \int_0^{2\pi}  \Sigma e^{- im \phi} d\phi \right| rdr, 
\end{eqnarray}
in the upper panel of Figure~\ref{fig:Fid-hydro-Cm}. The coefficients are normalized to the $m=0$ coefficient which corresponds to the total disk mass, $M_d$.

At early times, the disk surface density variations are dominated by the $m=1$ component. 
From the $C_1$ curve and the snapshots, this component appears as an exponentially growing, one-armed, trailing spiral with growth rate $\gamma_1 \approx 10^{-3}$.
Upon measuring the space-time evolution of the $m=1$ complex phase (shown in the bottom panel of Figure~\ref{fig:Fid-hydro-Cm}), we find that the spiral precesses {\it coherently} and in a prograde direction with a pattern speed of $\Omega_{\rm p} \approx 0.018$. 
Because $\Omega_{\rm p} \ll \Omega$ and the precession rate is coherent, what we are seeing is the growth of a slow eccentric mode \citep{Tremaine2001,Lin2015,Lee2019a}. 

During this time, higher $m$ modes are also exponentially growing with growth rates that are $\gamma_m \approx m \gamma_1$, albeit at an amplitude much less than $C_1$.
This relation suggests that these modes are harmonics of the $m=1$ mode and may be driven by non-linear, mode-mode interactions \citep{Laughlin1997,Laughlin1998,Lee2016}.

By $\sim 1700$ orbits, the growth stalls and $C_1$ saturates to a strength of $\sim 10\%$. 
This phase of evolution coincides with a morphological transition of the one-armed spiral into a set of closed elliptical rings and a slowdown in the $m=1$ pattern speed.

After $\sim 5000$ orbits, the eccentricity transitions yet again to being predominantly peaked in the inner regions of the disk ($r<R_0$). All modes start to decay from their peak amplitude on a timescale of $\sim 10^{4}/m$ orbits, and the rings transition to being mostly axisymmetric. While fading away, the $m=1$ component maintains a coherent pattern speed of $\Omega_{\rm p} \approx 10^{-2}$. 

\begin{figure}[t]
    \epsscale{1.1}
    \plotone{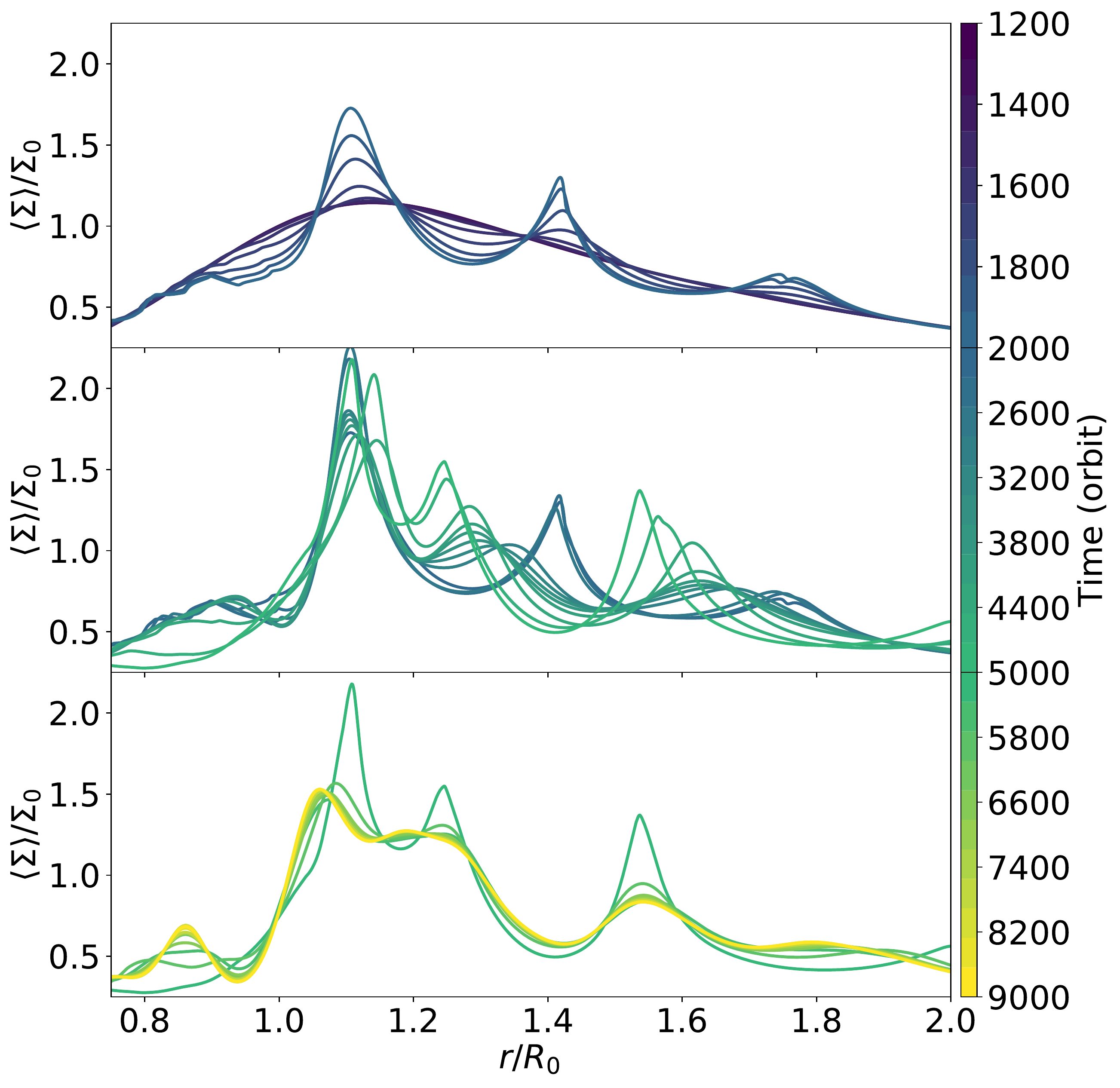}
    \caption{Time evolution of the azimuthally averaged density profile $\langle \Sigma \rangle$ in the  \textsc{Fiducial} simulation. 
    The three panels roughly correspond to the three phases of evolution shown in Figure \ref{fig:Fid-hydro-snapshot}: one-armed spiral (top), elliptical rings (middle), circular rings (bottom).
    }
    \label{fig:Fid-m0-t}
\end{figure}

\subsubsection{Ring evolution}

Figure~\ref{fig:Fid-m0-t} shows the azimuthally-averaged $\Sigma$ (denoted by $\avg{\Sigma}$) in the \textsc{Fiducial} simulation. 
During the growth phase (upper panel), $\avg{\Sigma}$ develops three well-defined over-densities that are at radii $r \approx 1.1, 1.5, 1.8 R_0$ and consistently grow over time. 
Once the instability saturates (middle panel), there is significant non-linear evolution of $\avg{\Sigma}$ that changes the ring locations.
In the final decay phase (bottom panel), $\avg{\Sigma}$ settles into a quasi-steady-state configuration consisting of three prominent rings at $r=1.1, 1.2, 1.5 R_0$, and additional, weaker features at $r = 0.85 R_0$ and $1.8 R_0$. 
Even though $\avg{\Sigma}$ increases in the rings, the disk maintains $Q \gtrsim 2$ throughout the simulation. 
We find no signatures of other instabilities, e.g. gravitational, Rossby wave, Rayleigh, or others.

With our \textsc{Fiducial} simulation, we have shown that a disk with inner and outer tapers traps an unstable one-arm spiral. 
This growing spiral arm has been previously reported by \citet{Lin2015} for locally isothermal disks ($\gamma=1$ and $\beta\rightarrow0$) and in cooled disks with $\beta \lesssim 0.1$, although the latter result was not discussed in detail and used preliminary simulations.
We have shown, in detail, that a non-zero cooling time unambiguously results in a similar instability.
But more importantly, we find that the non-linear outcome of this spiral instability is the generation of concentric axisymmetric rings.

%\amd{Viscous paragraph here.}
%\amd{(1) need $\nu k^2 < \gamma_1$. Prelim sims find $\nu < 10^{-5}$ gives same growth and spiral, which is $\alpha \lesssim 0.01$ at $1$.
%(2) condense next paragraph}

In reality, disk viscosity can affect the growth of the spiral and the lifetime of the rings. 
In our \textsc{Fiducial} simulation, viscosity is negligibly small. 
If we were to include a non-zero viscosity $\nu$, it could suppress the growth of the spiral if the damping rate is faster than the growth rate.
Moreover, a large $\nu$ may diffuse away the initial $\Sigma$ profile before any growth could occur.
Preliminary short-term simulations of the growth phase with constant viscosity find that $\nu \lesssim 10^{-5}$, which corresponds to $\alpha \lesssim 0.01$ at $R_0$, gives the same growth rate and spiral profile as our \textsc{Fiducial} results. 
We expect that in a long-term viscous simulation, the lifetime of the axisymmetric rings will be longer than the lifetime of the spiral waves as a tightly-wound spiral damps at a rate that is faster than the rings by a factor of $(k w)^2 > 1$, where $k$ is the radial wave number of the spiral and $w$ is the radial width of a typical ring.

In the next sections, we show from first principles why the disk is unstable and why that generically results in concentric rings.

\section{Theory of spiral instability and comparison to the fiducial results}  \label{sec:Linear} 

Our theory is based on the linearized equation of motion for eccentricity in an adiabatic disk with a finite cooling rate (Equation~\ref{eq:Fid-adi}). In Appendix~\ref{app:E-equation} we show that normal modes of the disk, e.g. $u \propto e^{-i \omega t + i m \phi}$, that are both slow ($\Re(\omega) \ll \Omega_K$) and eccentric ($m=1$) satisfy the eigenvalue equation, 

\begin{widetext} 
\begin{eqnarray}
    \label{eq:Linear-Ecool}
    2r^3\OK\Sigma \omega = && \underbrace{ 
    \frac{1}{1 + \beta^2} \frac{d}{dr} \left\{ r^3 P \left[ c_{\rm iso}^2 \frac{d}{dr} \left(\frac{E}{c_{\rm iso}^2}\right) + \gamma \beta^2 \frac{dE }{dr}\right] + i \beta r^3 P \left[ (\gamma-1) \frac{dE}{dr} + \frac{d \ln c_{\rm iso}^2}{dr} E \right] \right\}
    + r^2 \frac{dP}{dr} E
    }_{\text{pressure with $\beta$ cooling}}
    \nonumber \\ 
    && -
    \overbrace{
    \Sigma r\frac{d}{dr}\big(r^2\frac{d\Phi_0}{dr}\big)E - \Sigma \frac{d}{dr}\big(r^2\Phi_1\big)
    }^{\text{disk self-gravity}} %\nonumber
\end{eqnarray}
\end{widetext}
where $\omega$ is the complex mode frequency (the ``eigenvalue''), $E$ is the complex eccentricity (the ``eigenvector''), and $c_{\rm iso}^2 = P/\Sigma$ is the isothermal sound speed squared. The eccentricity equation includes the contribution from the $m=[0,1]$ self gravity potentials, $\Phi_{[0,1]}$, and the effects of cooling through $\beta$. 

Many previous authors have derived the governing equation of the disk eccentricity by adopting either the adiabatic \citep[e.g.,][]{Goodchild2006,Lee2019a,Lee2019b} or the locally isothermal approximation \citep[e.g.,][]{Teyssandier2016,Munoz2020}. Their results can be recovered from our Equation~\eqref{eq:Linear-Ecool} by taking the limits $\beta \to +\infty$ and $\beta \to 0$, respectively. 

We numerically solve the boundary-eigenvalue problem given by Equation~\eqref{eq:Linear-Ecool} for the complex eccentricity and mode frequency using a finite difference matrix method \citep[see e.g.,][]{Adams1989,Lee2019a}.
We use a uniformly log-spaced grid that covers $r \in(0.02,50)$ with $N=2048$ cells.
At the boundaries we set $dE/dr =0$, but note that since our density goes to zero at the boundaries, our choice of boundary condition has little effect on the modes. Solving the eccentricity equation gives a set of $N$ modes. We focus only on the mode which has the fastest growth rate as this one should be present in the hydrodynamical simulation.

\begin{figure}[t]
    \epsscale{1.1}
    \plotone{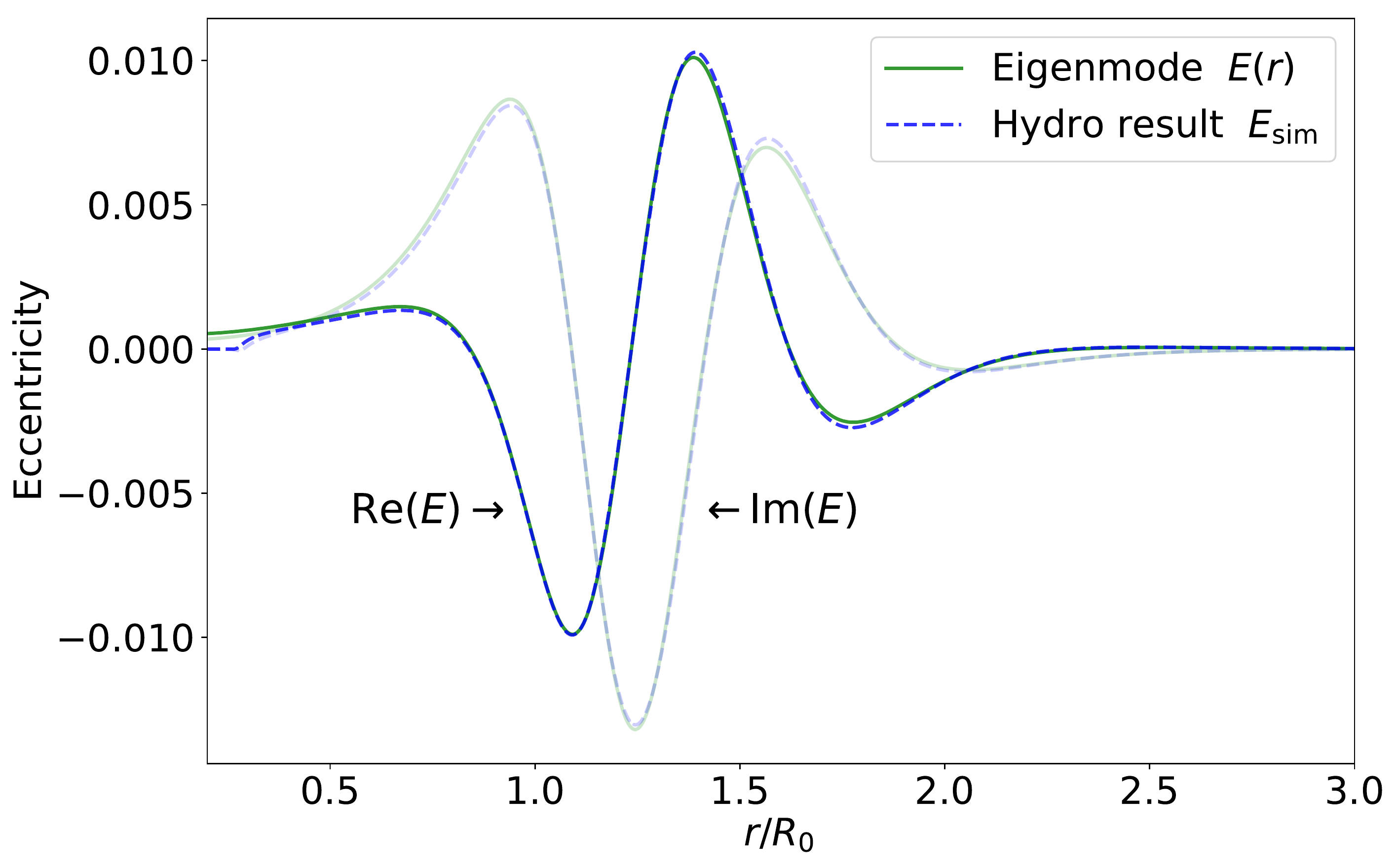}
    \caption{Radial profile of the (unstable) fundamental eccentric mode $E(r)$ in the \textsc{Fiducial} disk from the linear theory (Equation~\ref{eq:Linear-Ecool}; solid green) and the eccentricity $E_{\rm sim}$ from hydrodynamics simulation at $t=1500$ orbits (dashed blue). The real and imaginary parts are shown as opaque and transparent lines, respectively. The eigenfunction is multiplied by a real amplitude and a complex phase factor to fit the simulation result.
    }
    \label{fig:Linear-fid}
\end{figure}

Figure~\ref{fig:Linear-fid} compares the eigenfunction $E$ that we obtain for the \textsc{Fiducial} disk model to the eccentricity in the simulation $E_{\rm sim}$ (at 1500 orbits). The linear theory matches the hydrodynamics result almost perfectly.
The argument of periapse, $\arg(E)$, of the eccentricity varies with $r$, which produces the single-arm spiral that we see in the 2D gas density.

\subsection{Characteristics and growth of slow eccentric modes} \label{sec:Linear-analysis}

\begin{figure}[t]
    \epsscale{1.1}
    \plotone{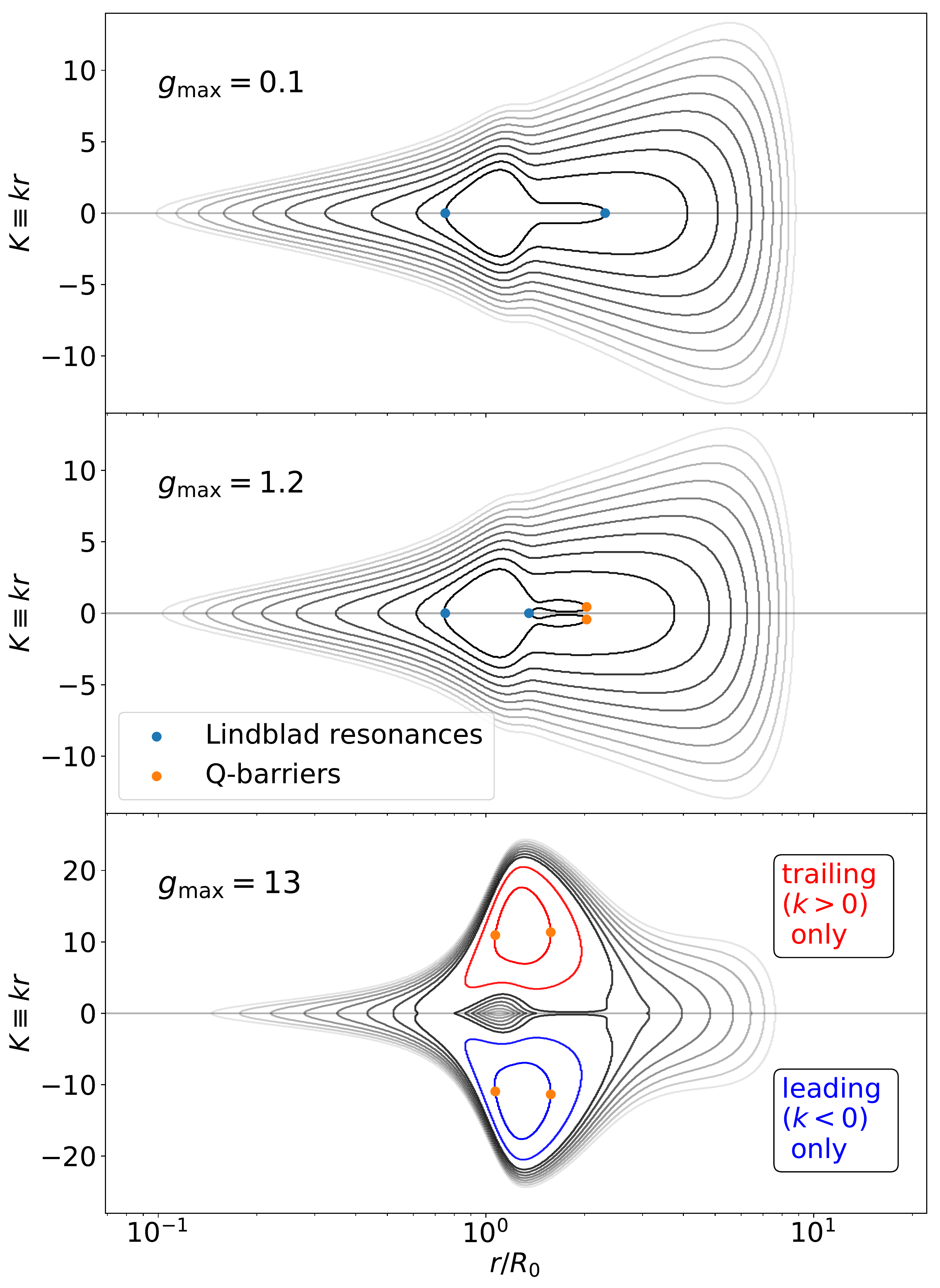}
    \caption{Dispersion relation map (DRM) for the eccentric modes in Equation~\eqref{eq:Linear-Ecool} based on the WKB dispersion relation derived in Appendix \ref{app:WKB}. The closed curves show the contours of constant $\omega$. Each panel adopts a different disk $\gmax$ (Equation~\ref{eq:Linear-g}). The darkest contour represents the fundamental mode, while the higher harmonics are displayed as more transparent contours. The elliptical modes are shown in black. The leading and trailing spiral modes are shown in blue and red, respectively. We also mark the Lindblad resonances (blue dots) and the Q-barriers (orange dots) for the fundamental mode defined by Equation \eqref{eq:app-wkb}.}
    \label{fig:Linear-DRM}
\end{figure}

Slow eccentric modes in self-gravitating, adiabatic disks have been thoroughly studied in \cite{Lee2019a}. They found that there are three kinds of modes that are different in morphology and controlled by the profile of 
\begin{eqnarray}
\label{eq:Linear-g}
g(r) = \frac{\pi G \Sigma r}{c_{\rm iso}^2},
\end{eqnarray}
which measures the relative strength of pressure and self-gravity\footnote{In disks with cooling, the sound-speed used in the definition of $g$ is a mix of the isothermal and adiabatic values controlled by $\beta$ (see e.g., Equation \eqref{eq:app-wkb} in Appendix \ref{app:WKB}). Since we mainly focus on rapidly cooled disks, we use the isothermal sound speed in $g$.}. 

We show a collection of the different mode types in dispersion relation maps (DRM) in Figure~\ref{fig:Linear-DRM}. 
The DRM plots show contours of fixed $\Re(\omega)$ in $r$-$kr$ space, where $k$ is the radial wave number of the mode.
The contours are calculated from a WKB-like approximation to the eccentricity equation (see Appendix~\ref{app:WKB} and \citealt{Lee2019a} for more details).
Following the naming convention of \citet{Lee2019a} the three mode types are:
\begin{itemize}
    \item Pressure dominated elliptical modes (e-p modes) occur when $g < 1$. These are long wavelength retrograde ($\Re(\omega)<0$) modes that are trapped by two inner Lindblad resonances (locations were $k=0$).
    \item Self-gravity dominated elliptical modes (e-pg modes) occur when $g > 1$ and switch between long and short radial wavelength at Q-barriers (locations where $kr = g$).
    \item Spiral modes (s-modes) occur when $g \gg 1$ and consist of leading ($k<0$) and trailing ($k>0$) spirals trapped by Q-barriers at short wavelengths. Each of these contours are separate, i.e, there are no Lindblad resonances connecting them.
    The mode seen in the \textsc{Fiducial} simulation is a trailing s-mode.
\end{itemize}

To determine whether a given mode is unstable we look at the equation governing the modes angular momentum deficit, $\Sigma r^3 \Omega |E|^2$, which can be obtained from multiplying the eccentricity equation by complex conjugate of $E$, $E^*$, and integrating in $r$ over the whole disk. The growth rate can then be determined from the imaginary component of this equation. We show in Appendix~\ref{app:growth-rate} that doing so results in two terms which contribute to the growth rate $\gamma_1=\Im(\omega)$:
\be
\label{eq:Linear-AMDintegral}
\gamma_1 \int 2 r^3 \Sigma \Omega |E|^2 dr = &&
-\int r^3 P \big[ \beta (\gamma-1) |\frac{dE}{dr}|^2
 \\
&& - \frac{d \ln c_{\rm iso}^2}{d r} \Im\left\{ (1-i\beta)  E\frac{dE^*}{dr} \right\} \big] dr,  \nonumber
\ee
to first order in $\beta$. For an elliptical mode which has $E \propto \cos(kr)$, the integrand on the right-hand side is proportional to $-k^2\cos^2(kr)$, so that there is no growth ($\gamma_1\leq0$), hence {\it elliptical modes are stable}. 

For an s-mode, $E \propto e^{i kr}$, so that $|dE/dr|^2=k^2|E|^2$ and $EdE^*/dr=-ik|E|^2$. From the DRM, we see that s-modes are radially confined, so that we can approximate the integral at the radius where $g=g_{\rm max}$. 
Doing so, we find that $\gamma_1$ is approximately,
\begin{eqnarray}
\label{eq:Linear-prediction-beta}
    \gamma_1 \approx -\frac{\gmax h^2 \Omega}{(2 \gamma)}[(\gamma-1) \beta\gmax + \frac{d \ln c_{\rm iso}^2}{d\ln r} {\rm sign}(k)],
\end{eqnarray}
where $h=c_{\rm s}/(r\Omega)$ and we approximate $kr\approx \gmax$. When $\beta$ is small, $\gamma_1$ is approximately linear to $\gmax$.

One can see that {\it only an s-mode can be unstable}. In particular, only modes with $k dc^2/dr < 0$ have the potential to be unstable, i.e., trailing (leading) spirals must be accompanied by decreasing (increasing) temperature profiles. 
For an instability to occur, $\beta$ must satisfy,
\be
\label{eq:Linear-beta_c}
\beta \lesssim 
\beta_{\rm c} \equiv
\frac{1}{1-\gamma} \frac{{\rm sign}(k)}{g_{\rm max} } \frac{d\ln c_{\rm iso}^2}{d \ln r}.
\ee 
Because slower cooling acts to suppress the instability, the disk must cool fast enough for an initial spiral mode to grow. 

\begin{figure}[t]
    \epsscale{1.1}
    \plotone{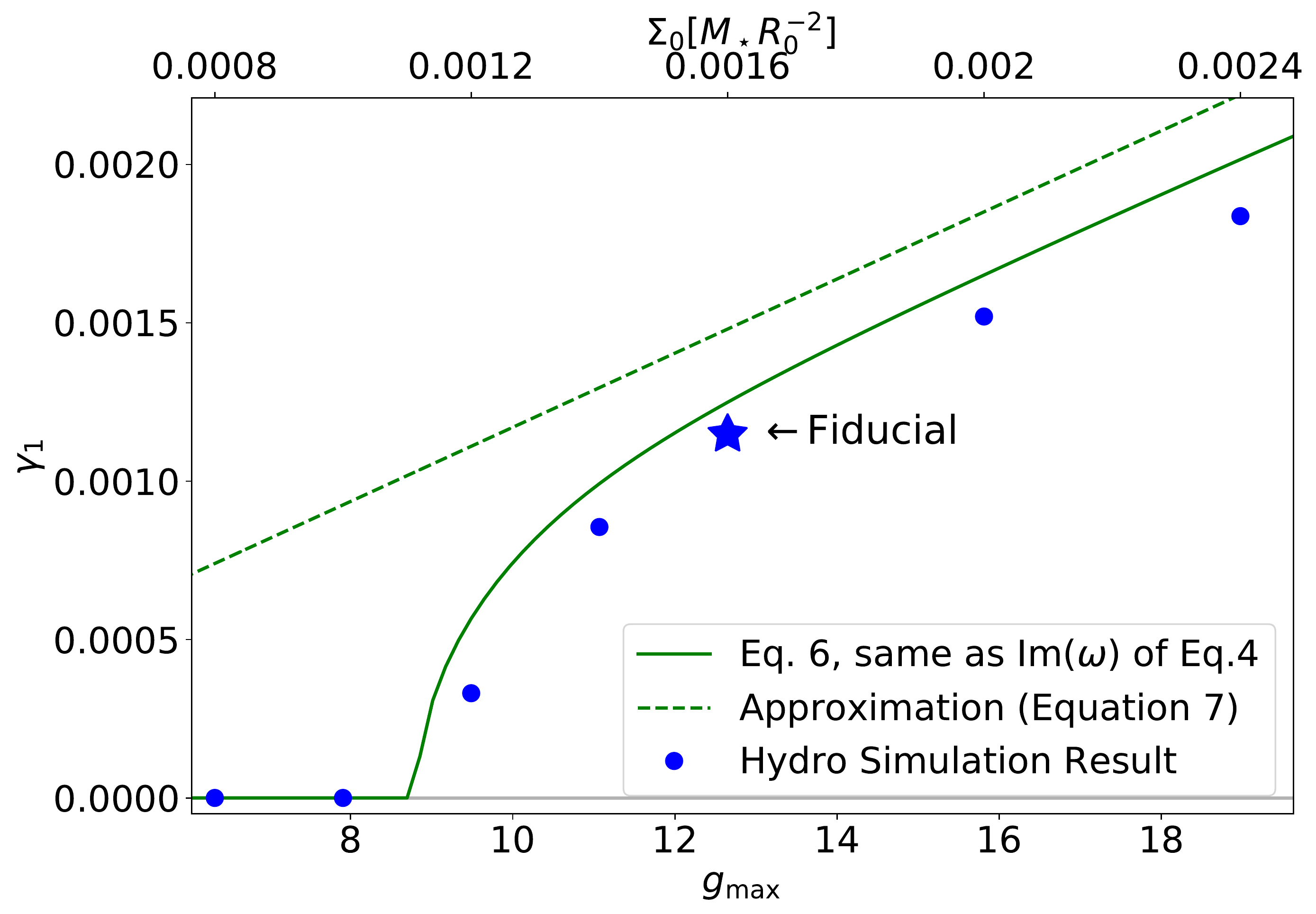}
    \epsscale{1.1}
    \plotone{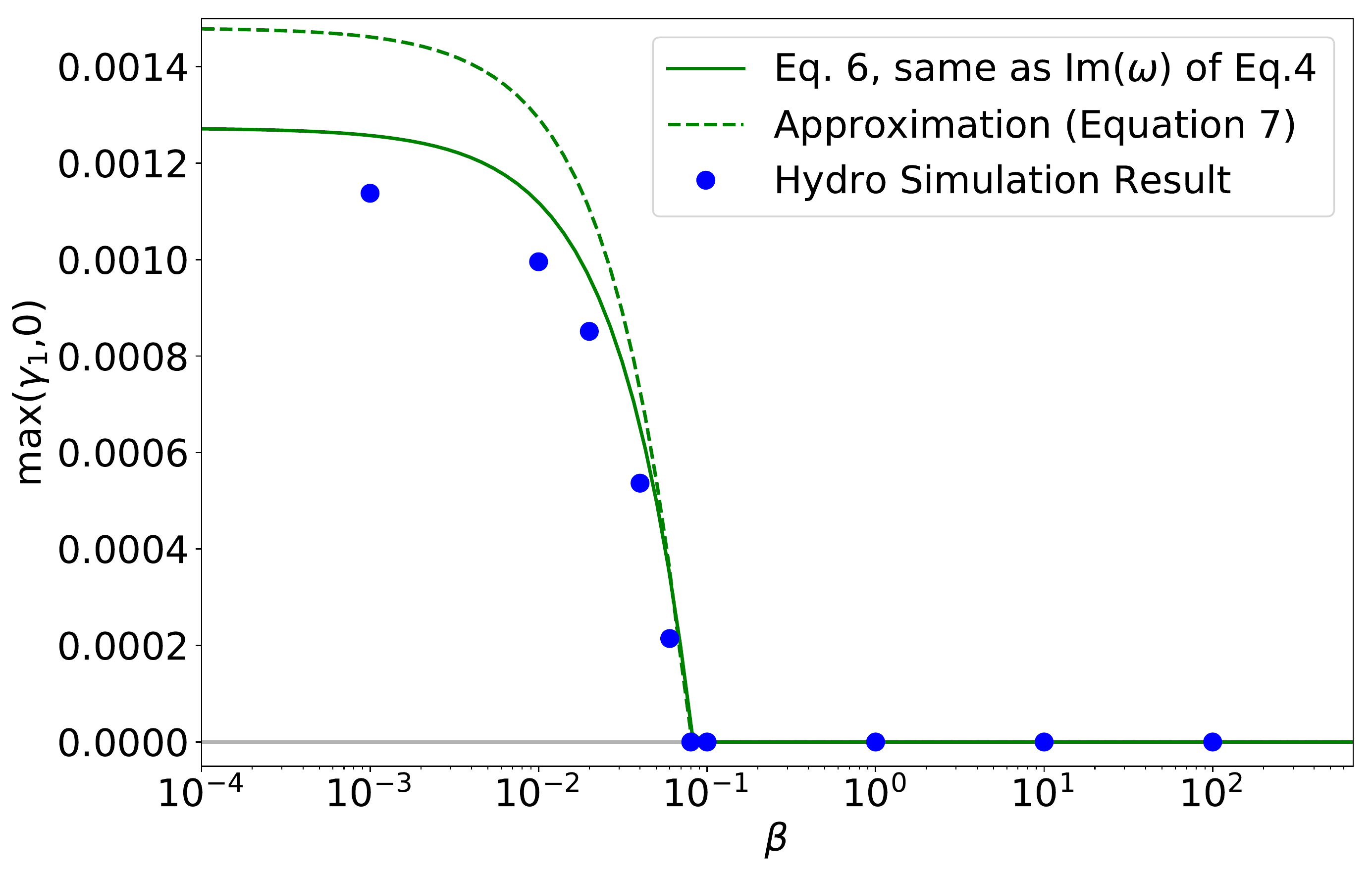}
    \caption{Growth rate of the fundamental eccentric mode as a function of $\gmax$ (top panel; all disks have $\beta=10^{-6}$) and $\beta$ (bottom panel; all disks have $g_{\rm max}=12.6$.
    The green line shows the rate calculated from the linear theory (Equation~\ref{eq:Linear-Ecool} and \ref{eq:Linear-AMDintegral}). The blue star and dots represent the rates measured from non-linear hydrodynamics simulations. The dashed green line shows the approximated prediction of $\gmax$ and $\beta$-dependence (Equation~\ref{eq:Linear-prediction-beta}). We plot max(0,$\gamma_1$) when $\beta>\beta_c$ because our simulations do not measure the suppression rate.
    }
    \label{fig:Linear-gr}
\end{figure}

Figure \ref{fig:Linear-gr} compares the growth rates between our linear results (solid curve) and a set of hydrodynamical simulations (blue dots) for different values of $g_{\rm max}$ and $\beta$. 
Both the linear and the hydrodynamics results suggest that an unstable mode exists when $\gmax>9$ (top panel). The linear theory growth rates agree to $10\%$ of hydrodynamics results for $\gmax>12$. Between $\gmax=9$ and $12$, the results differ by $71\%$ and $16\%$. This discrepancy is likely associated with the $\gmax$ being very close to the instability threshold. 
All of the larger $g_{\rm max}$ simulations remain gravitationally stable, $Q \gtrsim 2$, throughout their evolution.

For fast enough cooling rate (bottom panel), the growth rate $\gamma_1$ is almost a constant. 
For $\beta$ close to the critical value $\beta_{\rm c}$, the growth rate drops sharply from nearly maximum to zero as $\beta$ increases.
The hydrodynamic results suggest that $\beta_{\rm c}$ is between $0.06$ and $0.08$. Equation~\eqref{eq:Linear-AMDintegral} and Equation~\eqref{eq:Linear-beta_c} yields $\beta_{\rm c} = 0.08$ and $0.07$, respectively. 
The growth rates by Equation~\eqref{eq:Linear-AMDintegral}  agree within $13\%$ of the hydrodynamics results for $\beta\leq0.02$. 

Equation~\eqref{eq:Linear-prediction-beta} (dashed line) provides a good approximation to $\gamma_1$ for $\gmax>12$.
The main error introduced when going from Equation \eqref{eq:Linear-AMDintegral} to \eqref{eq:Linear-prediction-beta}, for $g_{\rm max} > 12$, comes from the approximation of the integral.
Despite this, Equation \eqref{eq:Linear-prediction-beta}, which does not require the $E(r)$ profile, only overestimates $\gamma_1$ by $\sim 10-20 \%$.
At smaller $g_{\rm max}$, Equation~\eqref{eq:Linear-prediction-beta} does not apply as the s-mode is either non-existent or not prominent.

In this section we have shown that an instability develops in a self-gravitating disk with cooling when (1) the ratio of self-gravity to pressure (i.e., $g$) is large enough to support a trapped spiral mode, (2) the spiral coincides with a non-zero temperature gradient, and (3) the cooling rate is faster than the critical rate given in Equation~\eqref{eq:Linear-beta_c}. Moreover, we have derived an analytical approximation for the resulting growth rate from the dispersion relation for the gas. In the limit of $\beta\rightarrow0$, our analytical growth rate agrees with the angular momentum based estimate of \cite{Lin2015}.

While we have focused on one particular choice of $\Sigma$ and $c_{\rm iso}^2$ profiles, the spiral instability mechanism applies to a broad range of astrophysically interesting density and temperature profiles. 
This is because the appearance of the s-modes is controlled by the radial profile of $g$, not $\Sigma$ or $c_{\rm iso}^2$ individually.
As we discuss in detail in Section \ref{sec:dis-profile}, this results in a wide range of density and temperature profiles that may be s-mode unstable.

\subsection{Ring formation in response to a growing spiral}  \label{sec:m0}

From Figure \ref{fig:Fid-m0-t}, we know that when there is a growing one-armed spiral, the disk also produces axisymmetric rings.
To understand this process, we examine how angular momentum is transferred from the spiral into the mean flow of the disk, and how this deposition process generically results in a non-constant radial mass flux that drives the formation of multiple rings.

In general, angular momentum is stored in the mean flow of the disk as well as any waves present. 
Angular momentum exchange occurs between these two reservoirs during wave excitation and when there is non-zero dissipation (from e.g., viscosity or shocks) or baroclincity (i.e., $\vec{\nabla} P \times \vec{\nabla} \Sigma \neq 0$).
These latter two processes are equivalent to the loss of conservation of the disk vortensity, $\xi = \hat{\vec{z}} \cdot (\vec{\nabla} \times \vec{v})/\Sigma$.

In inviscid disks with cooling, baroclinic effects transfer angular momentum from the mean flow, whose mass and angular momentum evolve as,
\be 
2 \pi r \pdt{\avg{\Sigma}} &+& \pdr{} (-\dot{M}) = 0 \label{eq:continuity} \\ 
2 \pi r^2 \pdt{\avg{\Sigma}\avg{v}} &+& \frac{\partial}{\partial r} \left( - \dot{M} r \avg{v}\right) = -t_{\rm dep} \label{eq:avg_ang}
\ee
into a wave, whose angular momentum follows\footnote{Note that $2 \pi r^2 \avg{\Sigma' v'}$ is the angular momentum in the wave in the Eulerian frame, i.e., at fixed radius. This is different than the wave angular momentum, $-2 \pi r^2\Sigma \Omega |E|^2 $, which can be calculated using the Lagrangian perturbations \citep{Lin2015}.} 
\be \label{eq:wave_ang}
2 \pi r^2 \pdt{\avg{\Sigma' v'}} +  \frac{\partial}{\partial r} \left(2 \pi r^2 \avg{\Sigma u v'} + F_G \right)= t_{\rm dep}
\ee
where $\dot{M}=-2\pi r \avg{\Sigma u}$ is the radial mass flux and $F_G$ is the flux of angular momentum due to self-gravity \citep{Goldreich1979}. 
The exchange torque density between the wave and mean flow is \citep{Dempsey2020}, 
\be \label{eq:tdep_def}
    \frac{t_{\rm dep}}{2\pi r} = -\avg{u' \Sigma'}\pdr{ (r \avg{v})} +\avg{\Sigma} \avg{u' \pdr{ (r v')}}  - \avg{\frac{\Sigma'}{\Sigma} \pdp{P}'}.
\ee
From the equation governing the evolution of the disk vortensity, $t_{\rm dep}$ for an eccentric wave can be shown to be \citep[][see also Appendix~\ref{app:tdep}]{Lubow1990},
\be
\label{eq:m0-tdep-E}
\frac{t_{\rm dep}}{4\pi r} = 
&-& \gamma_1 r^3 \Sigma^2 \pdr{\avg{\xi}} |E|^2  + r \pdr{c_{\rm iso}^2} \Sigma \Im\left(E^* r \pdr{E}\right) \nonumber \\ 
&+& \beta  (\gamma-1) r^2 c_{\rm iso}^2 \Sigma \left|\pdr{E}\right|^2,
\ee
in disks with $\beta \ll 1$.
The first term depends on the background vortensity gradient and is zero for stable modes, while the remaining terms are exactly the same as the terms on the right-hand-side of Equation \eqref{eq:Linear-AMDintegral} that set the growth rate.
We see now that they correspond to a non-zero, persistent background torque proportional to the disk's temperature gradient \citep{Lin2011,Lin2015}, and a stabilizing torque proportional to the cooling rate.

The loss of angular momentum from the mean flow induces a non-zero radial mass flux in the location of the wave.
From Equations~\eqref{eq:continuity}-\eqref{eq:avg_ang} this $\dot{M}$ is, 
\be
\label{eq:m0-Mdot-phys}
\dot{M} = \frac{2}{r \Omega_K} \left[2 \pi r^2 \avg{\Sigma} \pdt{\avg{v}} + t_{\rm dep} \right].
\ee
The $\partial_t \avg{v}$ term is mostly determined by the disk eccentricity, which is ultimately sculpted by $t_{\rm dep}$. 
We estimate $\partial_t \avg{v}$ by assuming the disk maintains an approximate radial force balance by modifying the pressure and self-gravity corrected Keplerian rotation, $\Omega_0$, by a small amount, $\Omega = \Omega_0 + \Omega_1$. 
From the azimuthally averaged radial velocity equation the $\Omega_1$ required to maintain balance is \citep{Lubow1990},
\be \label{eq:radial_force}
 2 r \Omega_0 \Omega_1 &=& \frac{1}{\avg{\Sigma}} \pdr{\avg{P}} + \pdr{\avg{\Phi}} -r \Omega_0^2  \nonumber \\
 &-& \avg{\frac{v'^2}{r}} + \avg{\frac{v'}{r} \pdp{u'}}  \nonumber \\
 &+& \avg{ \left(\frac{1}{\Sigma}\right)' \pdr{P'}} + \avg{u' \pdr{u'}} .
\ee
The first three terms on the right-hand side sum to zero as they constitute the dominate radial force balance. 
For slow eccentric modes the pressure term is $\sim h^2$ smaller than the velocity terms and so we can neglect it. 
Because we are interested in unstable modes, the remaining RHS terms grow as $e^{2 \gamma_1 t}$, so that the $\partial_t \avg{v}$ term in Equation~\eqref{eq:m0-Mdot-phys} is approximately,
\be 
\pdt{\avg{v}} &\approx& \frac{\gamma_1}{r \Omega_0} \left[  - \avg{ v'^2} +\avg{v' \pdp{u'}} + \avg{u' r \pdr{u'}} \right] \nonumber \\ 
&\approx& - \frac{1}{2} \gamma_1 r \OK |E|^2 +\gamma_1 r^2 \OK \pdr{}|E|^2 
\ee
where in the last line we replaced $u'$ and $v'$ with the complex eccentricity for an s-mode. 
The final induced $\dot{M}$ for rapid cooling is,
\begin{subequations}
\begin{align}
\dot{M} \approx
&+ 2\pi r^2 \Sigma \gamma_1 \left[ -|E|^2 + 2r\pdr{}|E|^2 \right] 
\label{eq:m0-Mdot-dldt_term}
\\
&- 8\pi r^3 \Sigma^2 \frac{\gamma_1}{\OK}\pdr{\avg{\xi}}|E|^2 
\label{eq:m0-Mdot-dxidr_term}
\\
&+ \frac{8\pi r}{\OK} \left[r \Sigma \frac{d c_{\rm iso}^2}{dr} \Im(E^*\pdr{}E) + \beta c_{\rm iso}^2 (\gamma-1) \left|\pdr{E}\right|^2 \right] .
\label{eq:m0-Mdot-T_term}
\end{align}
\end{subequations}
The terms proportional to $\gamma_1$ were previously derived by \citet{Lubow1990} for general $m$, while the last two terms are unique to a $\beta$-cooled disk with an eccentric spiral. 
To complete the calculation of $\dot{M}$ we can use the linear solution for $E$. One can then use Equation \eqref{eq:continuity} to compute the mean density evolution.

\begin{figure}[t]
    \epsscale{1.1}
    \plotone{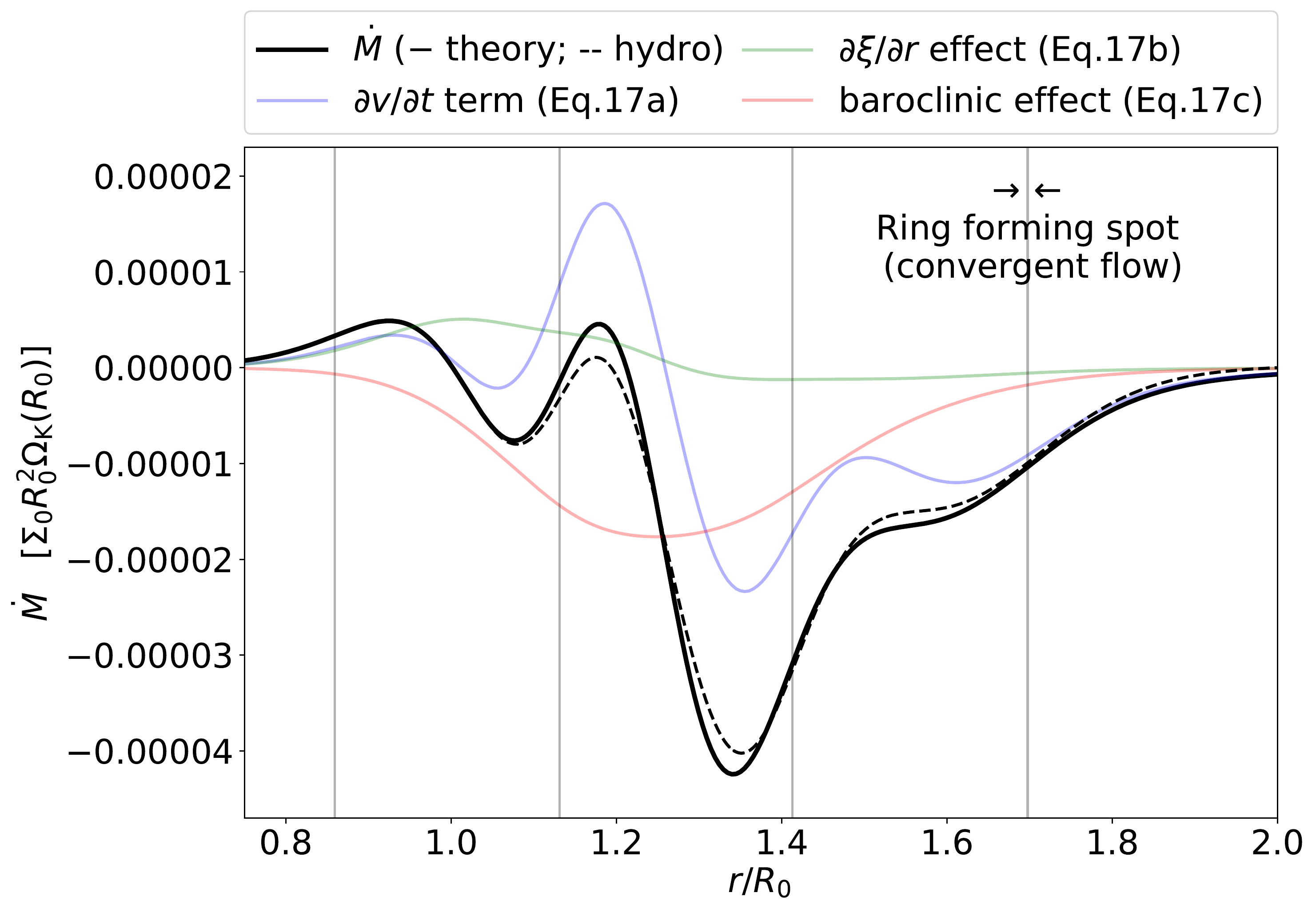}
    \caption{Radial mass flux, $\dot{M}$, at $1500$ orbits in our \textsc{Fiducial} simulation (dashed black) and theoretical predictions (solid lines) from Equation~\ref{eq:m0-Mdot-dldt_term}-\ref{eq:m0-Mdot-T_term}). 
    The colored curves show the contributions to $\dot{M}$ from the $\partial_t \avg{v}$ component (Equation~\ref{eq:m0-Mdot-dldt_term}), the $\partial_r \avg{\xi}$ component (Equation~\ref{eq:m0-Mdot-dxidr_term}), and the baroclinic component (Equation~\ref{eq:m0-Mdot-T_term}). 
    The $E(r)$ profile is calculated as in Section~\ref{sec:Linear} with the disk initial profiles and scaled to $|E|$ in the snapshot.
    In Equations \eqref{eq:m0-Mdot-dldt_term} - \eqref{eq:m0-Mdot-dxidr_term}, we use the measured $\gamma_1$ from the \textsc{Fiducial} simulation.
    The vertical lines represent the ring forming spots in the simulation. 
    }
    \label{fig:m0-rings}
\end{figure}

We show in Figure \ref{fig:m0-rings} that $\dot{M}$ as given by Equations~\eqref{eq:m0-Mdot-dldt_term}-\eqref{eq:m0-Mdot-T_term} agrees remarkably well with the $\dot{M}$ in the exponential growth phase of the \textsc{Fiducial} simulation.
In particular, we recover the approximate locations of the rings, which coincide with convergent points in the axisymmetric flow, i.e., at \textit{rising inflection points} of $\dot{M}$.
The primary ring locations are set by the $\partial_t \avg{v}$ terms in Equation \eqref{eq:m0-Mdot-dldt_term}. 
The vortensity (Equation \ref{eq:m0-Mdot-dxidr_term}) and baroclinic (Equation \ref{eq:m0-Mdot-T_term}) terms mostly shift the $\dot{M}$ profile and slightly modify the ring locations.
The small disagreement between the linear and hydrodynamical results is primarily due to the $m\neq1$ perturbations and the terms of $\mathcal{O}(\gamma_1h^2,\gamma_1\beta)$ that we have dropped.

Overall, this excellent agreement demonstrates that the initial mass redistribution leading to the formation of multiple rings occurs because of the exchange of angular momentum from the background disk and the spiral wave. 

\subsection{Post-saturation evolution}
\label{sec:Fid-sat}

While the initial phase is linear and fully describable by our linear theory, non-linear evolution takes over at later times (cf. Figure \ref{fig:Fid-m0-t}). In our \textsc{Fiducial} simulation, $\avg{\Sigma}$ eventually evolves to the point where the instability saturates at $t\sim1700$ orbits.
By this time, the disk can no longer trap the s-mode and instead the fundamental mode becomes elliptical. 

The ring formation driven by the s-mode also stops with the saturation of the instability. 
Follow-up non-linear evolution can still modify the sharpness and locations of the rings until the elliptical mode also fades away.
We suspect this is caused by the inner boundary condition in the \textsc{Fiducial} simulation, because an elliptical eccentric mode dominated by self-gravity, unlike an s-mode, tends to have larger eccentricity near the inner edge of the disk \citep{Lee2019a}. 
The lifetime of such a mode will thus be short as any dissipation -- whether physical or numerical -- will operate on a faster timescale than the dissipation of the rings at larger radii.

\section{Discussion} \label{sec:dis}

\subsection{Disk shape and temperature profile}
\label{sec:dis-profile}

In Section~\ref{sec:Linear-analysis}, we showed that disks are unstable when they can both cool faster than a critical rate and trap a slow spiral mode. 
Spiral modes appear when the distance (in phase-space) between two Q-barriers is large enough to satisfy a quantum condition \citep[][see also Figure~\ref{fig:Linear-DRM}]{Lee2019a}. 
Because Q-barriers occur where $k r = g(r)$, the $g$-profile needs to be both peaked at some radius and wide enough to provide sufficient separation of the Q-barriers.
This suggests that the larger scale properties of the $\Sigma$ and $c_{\rm iso}^2$ profiles are unimportant, and that only the local properties in the vicinity of the $g$ maximum are relevant.

Consider temperature and density profiles which produce a peak in $g$ at the radius $r_{\rm max}$. 
Taylor expanding $g(r)$ around this peak produces a simple Gaussian g-profile,
\be \label{eq:dis-gprof}
g(r) = g_{\rm max} e^{- (r-r_{\rm max})^2/\sigma_g^2} ,
\ee
where $\sigma_g = \sqrt{2 (d^2 \ln g / dr^2)^{-1}}$ is the Gaussian width. 
For a given $r_{\rm max}$, $g_{\rm max}$, and $\sigma_g$ the density and temperature profiles need only satisfy the following requirements at $r_{\rm max}$:
\be
g_{\rm max} &=& \left. \frac{ \pi G \Sigma r}{c_{\rm iso}^2} \right|_{r_{\rm max}} , \label{eq:gcond1} \\
\frac{d}{d\ln r }  \ln \left(\frac{r \Sigma}{c_{\rm iso}^2} \right)_{r_{\rm max}} &=& 0 , \qquad 
\sigma_g^{-2} = \frac{1}{2}  \frac{d^2}{d r^2}\ln \left(\frac{c_{\rm iso}^2}{\Sigma} \right)_{r_{\rm max}} . \label{eq:gcond2}
\ee
A wide range of $\Sigma$ and $c_{\rm iso}^2$ profiles can be constructed to fulfill these requirements. 

\begin{figure}[t]
    \epsscale{1.1}
    \plotone{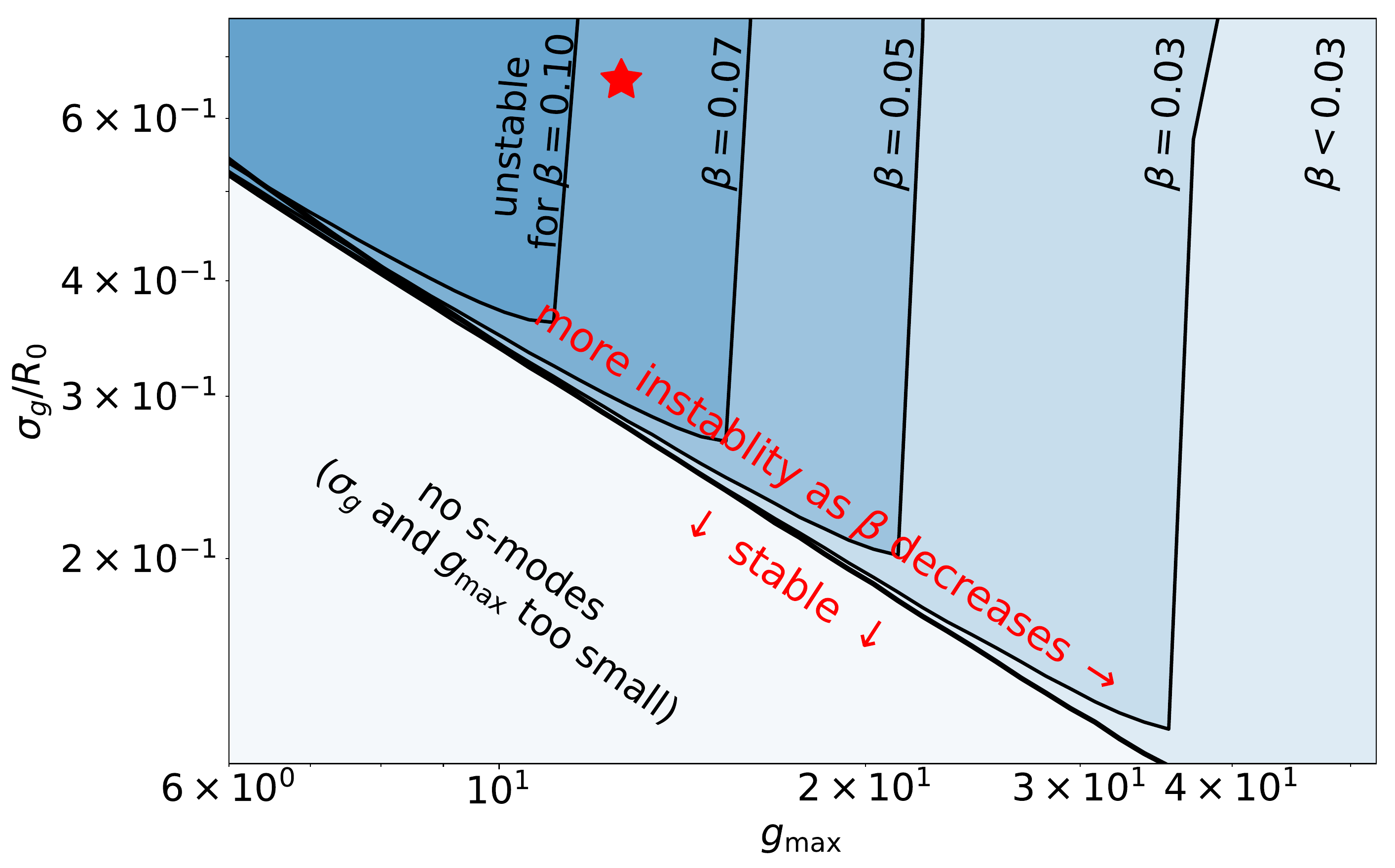}
    \caption{Stability of the eccentric mode for different combinations of $\gmax$, $\sigma_g$, and $\beta$. 
    Bluer regions correspond to faster growth rates, while white regions are stable.
    Spiral modes may only exist above the solid line $\gmax \sigma_g/R_0 \sim 3$. 
    The black curves mark the lower boundaries of the unstable region ($\gamma_1 = 0$ contour) for the cooling rate $\beta$ as specified. The unstable region expands as $\beta$ decreases. The curves are calculated from the linear equation (Equation~\ref{eq:Linear-Ecool}) for $g$ as Equation~\eqref{eq:dis-gprof}. The red star represents the $g$ profile of the \textsc{Fiducial} disk.
    }
    \label{fig:dis-gmax-vs-gwid}
\end{figure}

Using this $g(r)$ profile, we solve for the linear $\gamma_1$ in the $(g_{\rm max}, \sigma_g, \beta)$ parameter space.
Our results are shown in Figure \ref{fig:dis-gmax-vs-gwid} for $c_{\rm iso}^2 \propto r^{-1/2}$ and $\Sigma$ given by Equations \eqref{eq:Linear-g} and \eqref{eq:dis-gprof}. 
We find that there is a region of $\sigma_g$-$g_{\rm max}$ parameter space that is unstable to one-armed spirals. 
This region is bounded at large $g_{\rm max}$ by Equation \eqref{eq:Linear-beta_c} where the cooling is too slow, and below by $g_{\rm max} \sigma_g / R_0 \sim 3$, where the $g$-profile is too narrow to trap spiral modes. 
Lowering $\beta$ extends the right boundary of the unstable region to higher $g_{\rm max}$ values. 
Using Figure \ref{fig:dis-gmax-vs-gwid} and Equations \eqref{eq:gcond1}-\eqref{eq:gcond2} we can thus estimate whether any given $\Sigma$ and $c_{\rm iso}^2$ profiles will become unstable. 

\begin{figure}[t]
    \epsscale{1.1}
    \plotone{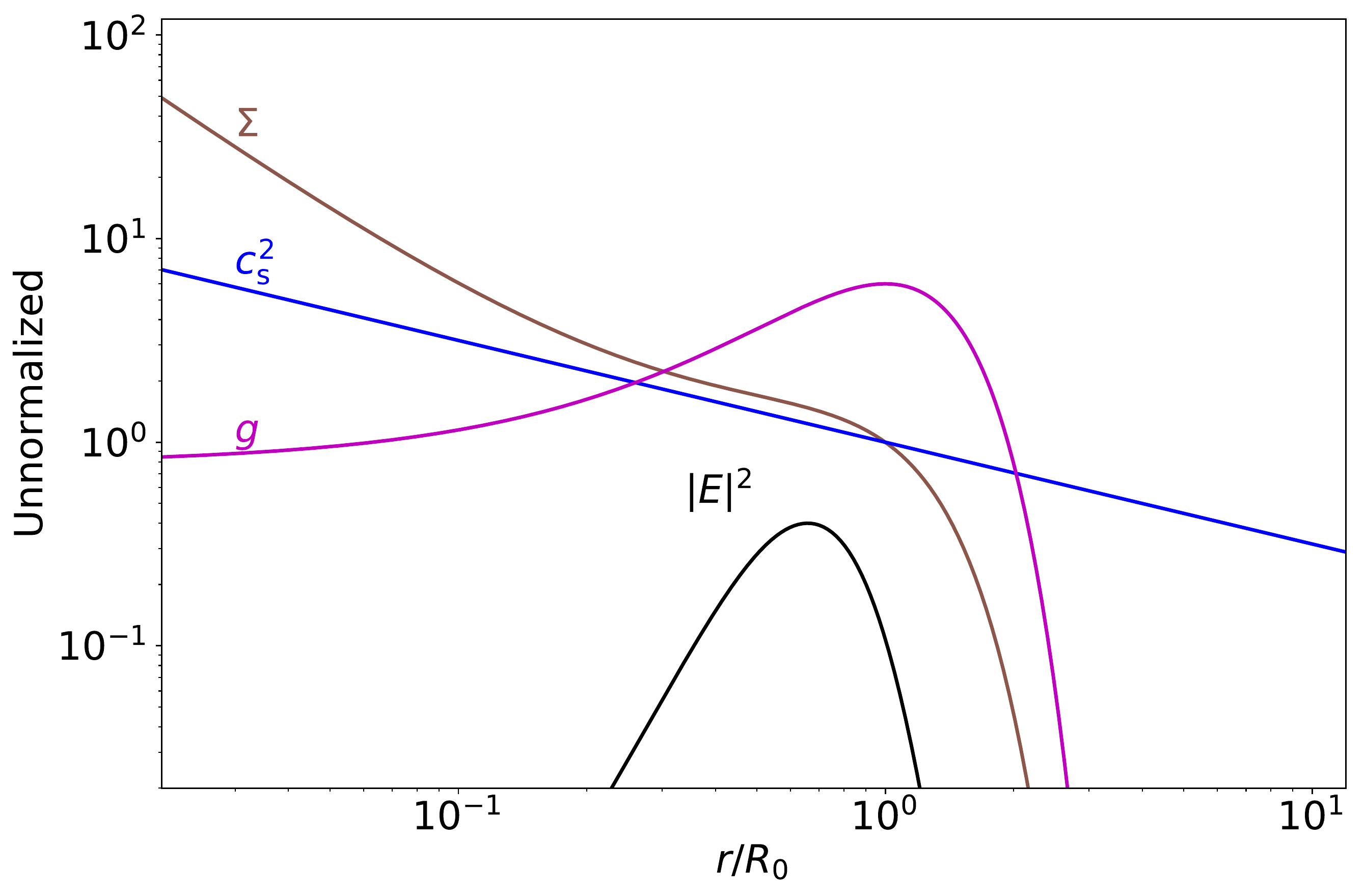}
    \caption{Example of a disk with a monotonic density, power-law sound speed, and spiral-trapping $g$ profile with $\gmax=9$, $\sigma_g=0.7R_0$, $r_{\rm max}=R_0$. The black curve shows the unnormalized linear $|E|^2$ profile.}
    \label{fig:dis-profile}
\end{figure}

One limitation of our \textsc{Fiducial} simulation is that our $\Sigma$ profile has both an inner hole and an outer taper. 
Such a model may be representative of observed transition disks, which are characterized by large inner cavities \citep[e.g.,][]{Muzerolle2010,Espaillat2014,Ansdell2016,Ansdell2017}.
However, transition disks make up only $\lesssim 20\%$ of observed protoplanetary disks.
In Figure \ref{fig:dis-profile}, we show that
{\it monotonic density profiles}, i.e. disks without inner holes, can also produce peaked $g(r)$ profiles capable of trapping unstable s-modes.
Hence, the s-mode instability may be a generic instability capable of producing multiple rings across a variety of observed protoplanetary disks.

Our results are independent of what physical length units we give to $R_0$.
Thus, we can take the $\Sigma$ profiles shown in Figures \ref{fig:Fid-profile} and \ref{fig:dis-profile} and apply them to realistic disks.
Recall that the s-mode instability requires both a large $g_{\rm max}$ and small $\beta$. 
For a given disk mass we can place a constraint on what $H/r$ is required to produce a $g_{\rm max} \gtrsim 10$.
For example, a transition disk (similar to Figure \ref{fig:Fid-profile}) with a cavity inside of $50$ AU and disk mass $\sim 0.05 M_\odot$ would need $H/r \lesssim 0.06$ at $50$ AU to attain $g_{\rm max} \gtrsim 10$. 
Similarly, a non-transition disk (similar to Figure \ref{fig:dis-profile}) with an exponential cutoff outside $150$ AU at the same mass would require $H/r \lesssim 0.05$ at $150$ AU to attain $g_{\rm max}$. 
To see whether such disks will actually be unstable requires an estimate of the cooling rate, which we focus on in the next subsection.

\subsection{Cooling rate}
\label{sec:dis-beta}

So far, we have focused exclusively on cooling timescales with constant $\beta$.
In contrast, more realistic, radiative cooling that takes into account various opacity sources in the disk produces a non-constant $\beta$ profile.
A more general $\beta$ function affects both the ability of the disk to trap spiral modes, through, the g-profile, as well as the criterion for instability (Equation \ref{eq:Linear-beta_c}). 

From \citet{Zhang2020}, we can assign an effective $\beta$ to a more realistic radiative cooling model,
\begin{eqnarray}
    \beta = 
    & 0.015 & (\frac{f}{0.01})^{-1} (\frac{\kappa_{R}}{1{\rm cm}^2/{\rm g}})^{-1}(\frac{L_{\star}}{L_{\sun}})^{1/2} (\frac{\phi}{0.02})^{-3/4} 
    \nonumber\\
    &\times& (\frac{\Ms}{M_{\sun}})^{1/2}(1+\tau^2),
\end{eqnarray}
which depends on the central star properties ($M_\star,L_\star$), the dust-to-gas ratio $f$, the flaring angle, $\phi$, the disk opacity, $\kappa$, and the optical depth, $\tau$. 
For typical protoplanetary disk parameters, $\beta$ can easily be $\ll 1$, i.e, realistic cooling may destabilize a disk with a trapped spiral mode. 
Even smaller $\beta$ can be achieved by raising the dust-to-gas ratio or the opacity, or by looking at disks around less massive or less luminous stars. 

The above estimation does not take into account the effects of non-constant $\beta$ on the disk stability. 
In Appendix~\ref{app:E-equation} we derive the correction to the disk normal mode equation (Equation~\ref{eq:app-E-equation-Mbeta}) for a radius-dependent $\beta$. 
To the lowest order in $d\beta/dr$, there is an extra torque on a slow eccentric mode proportional to $(d\beta/dr)(dc^2/dr) \Sigma |E|^2 $.
This torque modifies the criterion for instability to, 
\be
\beta < \beta_c = \frac{1}{g(\gamma-1)} \left(- \frac{d\ln c^2}{d\ln r} \right) \left[ {\rm sign}(k) - \frac{1}{g} r\frac{d \beta}{dr} \right] . 
\ee
Depending on the sign of $d\beta/dr$, this extra term can either help stabilize or further destabilize a spiral-mode. 
Note that this additional torque is independent of the mode type, i.e., elliptical modes also experience a torque proporitional to $(d\beta/dr) (dc^2/dr)$. 
This raises the intriguing possibility that elliptical modes may also be unstable if $(d\beta/dr) (dc^2/dr) < 0$.

\subsection{3D effects} \label{sec:dis-3d}

In this work, we have only considered 2D disks. 
However, the 3D simulations in \citet{Lin2015} mostly confirmed the 2D results.
He found that the s-mode instability can still occur in 3D disks, but with a smaller growth rate. 
The spiral pattern varies with height, while the mid-plane pattern is similar to the 2D results. 

For the linear theory, an additional pressure-like term on the order of $~\sim h^2 E$ should be added to Equation \eqref{eq:Linear-Ecool} in 3D \citep{Ogilvie2008}.
Hence, the $\gmax$ and $\beta_c$ criteria for the spiral instability may change. 
But, our ring generation mechanism is derived for an arbitrary $m=1$ perturbation until we plug in the eccentricity formalism in the last step, thus we expect the ring formation to occur in 3D disks as well. 
Nevertheless, long-term 3D simulations still need to be done in the future to explore the important non-linear outcomes and final ring configuration.

\subsection{Implications for observations}  \label{sec:dis-obs}

Our model manifests a variety of disk substructures that may be observable in dust continuum images. 
To get a sense of the dust response, we have performed a locally isothermal simulation with dust \citep[see e.g.,][for a description of the dust implementation in \texttt{LA-COMPASS}]{Fu2014}. 
The dust is initialized to the same distribution as the gas with a dust-to-gas ratio of $0.01$ and is not allowed to feedback onto the gas. 
For simplicity, we only consider dust with a diameter of 0.2mm and internal density of 1.25g/cm$^3$. 
By assuming a solar-type star and a reference radius $R_0=20$AU, the 0.2mm grains have a Stokes number of $\sim 10^{-3}$ at $R_0$. 

\begin{figure}[t]
    \epsscale{0.847}
    \plotone{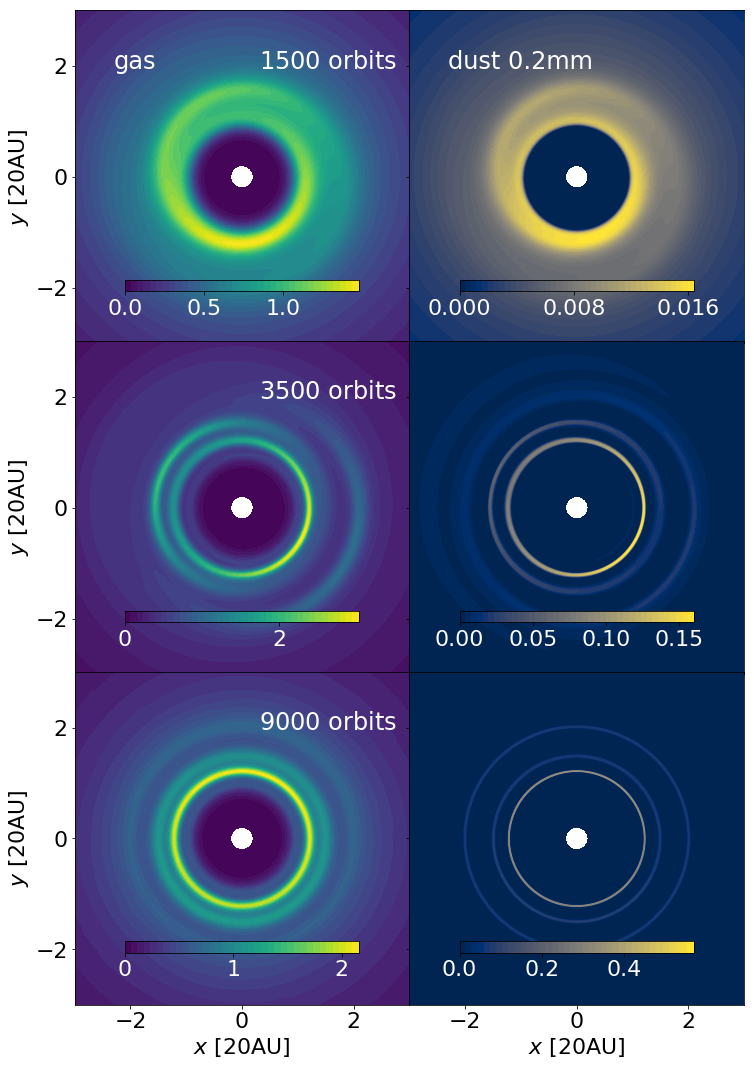}
    \epsscale{0.2695}
    \plotone{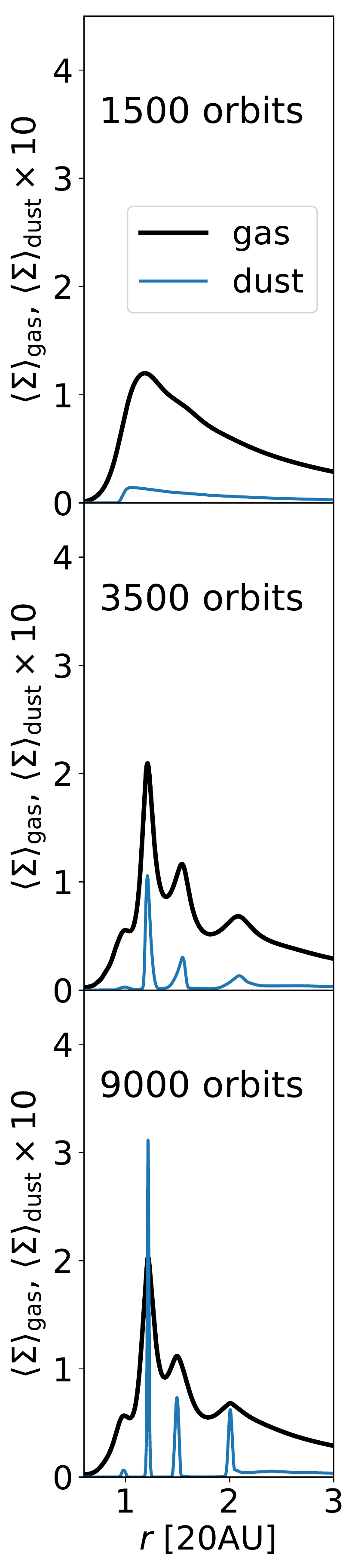}
    \caption{2D Snapshots of the hydrodynamics simulation with dust (left) and $\avg{\Sigma}$ (right) at time $=1500$ orbits (top), $3500$ orbits (middle), and $9000$ orbits (bottom). 
    The simulation includes 0.2mm dust initialized to the same distribution as the gas with a 0.01 dust-to-gas ratio. The Stokes number at $R_0$ is $\sim 10^{-3}$ initially. 
    The unit of the colorbars in the left panel and the vertical axes in the right panel is $\Sigma_0$.}
    \label{fig:obs-dust-snapshot}
\end{figure}

Figure~\ref{fig:obs-dust-snapshot} shows that the dust largely resembles the gas pattern. 
The $\avg{\Sigma}$ plots demonstrate that the primary dust ring is formed at the primary gas ring ($r\sim1.2R_0$). 
The two comparable dust rings coincide with the secondary and tertiary gas ring ($r\sim1.5, 2.3R_0$). 
We expect the dust rings to be wider in viscous disks. 
This result suggests that our substructure formation mechanism may create significant observational features in the dust continuum. 
Further work including a range of dust sizes and dust-feedback is required to make more realistic observational predictions.

Kinematical signatures of disk eccentricity may also be observable.
During the spiral instability stage in our model, the eccentric motions of the gas reach $\sim$ 10\% of the Keplerian motion, and in the final stage, the rings may still possess some residual eccentricity. 
Gas kinematics traced by any molecular line emission can verify whether or not the mechanism we present in this paper is the true origin of some observed disk substructures. 
A caveat here is that we have considered only two-dimensional disks. As we discussed in Section~\ref{sec:dis-3d}, while the starlight is supposedly scattered at the top and bottom surfaces of the disk, it is not clear if the disk eccentricity should be independent of height. 

Observations of a spiral instability in disks may also provide new constraints to the mass, opacity, and other basic disk parameters because the instability depends on the disk properties such as the self-gravity-to-pressure ratio and cooling efficiency.

\section{Summary}  \label{sec:Summary}

In this paper, we have explored the properties and long-term evolution of a spiral instability in 2D self-gravitating (but gravitationally stable) disks with cooling.
Our main results can be summarized as follows:

\begin{enumerate}

    \item With our \textsc{Fiducial} long-term hydrodynamics simulation, we demonstrated that there is a trapped one-armed spiral instability that evolves into a set of axisymmetric rings over long timescales when the disk cools rapidly.

    \item We presented a linear equation that governs the slow global eccentric modes. We showed that the disk must have large enough $\gmax$ and must cool faster than $\beta_{\rm c} \sim \gmax^{-1}$ for the instability to occur. 
    The eigensolutions we obtain from our linear theory can accurately describe our non-linear hydrodynamics simulations in terms of the mode morphology, growth rate, and initial ring locations.

    \item We showed that the ring-forming mass flux relies on a persistent angular momentum transfer between the axisymmetric background and the non-axisymmetric wave. Furthermore, we presented a first-principle method for estimating the locations and amplitudes of the rings using only the initial equilibrium of the disk and  the eccentric mode from our linear theory.
    
    \item We showed that the s-mode instability does not rely on a specific model of density or temperature profile. As long as the disk profile of $g$ exhibits s-mode trapping regions, the spiral and subsequent ring forming process can be expected to be generic.   
    
\end{enumerate}

To conclude, our results demonstrate a new mechanism to form disk substructures {\it without planets}. 
The morphology and kinematics of these substructures can be quantitatively predicted by our analytical theory. 
Our theory may also provide new interpretation to the protoplanetary disk observations.

\acknowledgments
We thank the anonymous reviewer for valuable comments and suggestions.
Helpful discussions with Yaping Li are gratefully acknowledged.
JL thanks Dong Lai for helpful advice.
This work is supported by the Laboratory Directed Research and Development Program (LDRD) and the Center for
Space and Earth Science at Los Alamos National Laboratory (approved for public release as LA-UR-20-27570)
as well as by an NASA/ATP project.  

\vspace{15mm}

\software{LA-COMPASS \citep{Li2005,Li2008},  
          Matplotlib \citep{Hunter2007}, 
          NumPy \citep{Walt2011},
          SciPy \citep{Virtanen2020}
          }

%\clearpage
%\newpage
\vspace{2cm}

\appendix

\section{Eccentric normal mode equation}
\label{app:E-equation}

We start with re-writing the energy equation 
(Equation \ref{eq:Fid-adi}) in terms of the gas entropy, $S \equiv \ln(P/\Sigma^{\gamma})$, 
\begin{equation}
\label{eq:app-E-equation-SE}
    \Sigma T (\frac{\pd S}{\pd t} + u\frac{\pd S}{\pd r} + \frac{v}{r}\pdp{S}) = - \Sigma \frac{T - T_{\rm eq}}{t_c},
\end{equation}

Now we assume that the background state is steady and axisymmetric, while the perturbation is small and has $m=1$ symmetry. After removing the equilibrium background from Equation~(\ref{eq:app-E-equation-SE}) and linearizing, we get
\begin{equation}
    -i(\omega-\Omega) \Big[\frac{P'}{P} - \gamma \frac{\Sigma'}{\Sigma}\Big] + u'\Big[\frac{1}{P}\frac{d P}{d r} - \gamma \frac{1}{\Sigma}\frac{d \Sigma}{d r}\Big] = \frac{\Sigma' T - P'}{\tc P},
\end{equation}
where $\omega$ is the mode frequency and we have approximated $T_{\rm eq} = T$. Unprimed quantities denote the axisymmetric background, while primed quantities are the $m=1$ perturbations.

Using the slow-mode approximation $\Omega=\OK=\sqrt{G\Ms/r^3}\gg|\omega|$ and defining $\beta \equiv \OK t_c$, the pressure perturbation $P'$ can be expressed as
\begin{equation}
\label{eq:app-E-equation-Pcool}
\begin{split}
    P' & = \frac{i\beta}{i\beta + 1}
    \underbrace{
    \left(-\gamma r P \frac{d E}{d r} - r \frac{d P}{d r}E\right)
    }_\text{adiabatic pressure}
    + \frac{1}{i\beta + 1}
    \underbrace{
    \left(-r P \frac{d E}{d r} - \frac{rP}{\Sigma}\frac{d\Sigma}{d r}E\right)
    }_\text{isothermal pressure},
\end{split}
\end{equation}
where we have expressed the perturbations using the complex eccentricity, $E$, defined as \citep{Lee2019a}
\begin{eqnarray}
\label{eq:app-E-equation-uvS-E}
u' = i\Omega r E, \quad v' = -\frac{1}{2}\Omega r E \quad \text{and} \quad \Sigma' = -r\frac{d}{dr}(\Sigma E).
\end{eqnarray}
Note that $P'$ is a linear combination of the pressure perturbations in the adiabatic limit and the locally isothermal limit. 

Following \citet{Lee2019a}, using $P'$ and the linearized equations of motion for $u'$, $v'$, and $\Sigma'$, we arrive at the normal mode eccentricity equation with cooling,
\begin{equation}
    \label{eq:app:E-equation-Ecool}
    2r^3\OK\Sigma \omega E = [\frac{i\beta}{i\beta+1} M_{\rm adi} + \frac{1}{i\beta+1} M_{\rm iso} + M_{\rm dsg} + M_{\beta}]E.
\end{equation}
where the four operators are,
\be
    M_{\rm adi}E &=&\frac{d}{dr}\left(\gamma r^3 P \frac{dE}{dr}\right) + r^2 \frac{dP}{dr}E,    \\
    M_{\rm iso}E &=&  \frac{d}{dr}\left(r^3 P \frac{dE}{dr}\right) + r^2 \frac{dP}{dr}E  - \frac{d}{dr}\left(\Sigma\frac{d c_{\rm iso}^2}{dr} r^3E\right) ,\\
    M_{\rm dsg}E &=& -\Sigma r\frac{d}{dr}\left(r^2\frac{d\Phi}{dr}\right)E - \Sigma \frac{d}{dr}(r^2\Phi_1) ,\\ 
    M_{\beta} E &=& P r^2 \left[\frac{d}{dr}\left(\frac{\beta^2 + i\beta}{1 + \beta^2}\right)\left(\gamma r  \frac{d E}{d r} + \frac{r}{P} \frac{dP}{dr}E\right)  + \frac{d}{dr}\left(\frac{1-i\beta}{1 + \beta^2} \right)\left(r \frac{d E}{d r} + \frac{r}{\Sigma}\frac{d\Sigma}{dr}E\right) \right] .   \label{eq:app-E-equation-Mbeta}
\ee
These quantify the adiabatic, isothermal, self-gravity, and non-constant $\beta$ effects.
Expressions for $\Phi_0$ and $\Phi_1$ can be found in Appendix A.3 of \citet[][]{Lee2019a}. 
Setting $M_\beta=0$ and simplifying results in Equation \eqref{eq:Linear-Ecool}.

\section{WKB dispersion relation} \label{app:WKB}

Following \citet{Lee2019a,Lee2019b}, the dispersion relation governing an eccentric wave can be derived using a WKB-like approximation. 
By letting $y=(r^3P)^{1/2}E$, one can transform the pressure operators into 
\be 
    M_{\rm adi}E &=& \gamma (r^3P)^{1/2} \frac{d^2}{dr^2} y + \left[ \sqrt{\frac{r}{P}} \frac{d P}{dr} - \gamma \frac{d^2}{d r^2}(r^3P)^{1/2} \right] y, \\
    M_{\rm iso}E &=&  (r^3Pc_{\rm iso}^2)^{1/2} \frac{d^2}{dr^2} \left(\frac{y}{c_{\rm iso}}\right) + \left[ \sqrt{\frac{r}{P}} \frac{d  P}{dr}  - \frac{1}{c_{\rm iso}}\frac{d^2}{dr^2}(r^3Pc_{\rm iso}^2)^{1/2} \right] y.
\ee 

For the self-gravity operator, we use the same second-order expression as \citet{Lee2019a}, and set $M_\beta=0$.
Transforming $E \rightarrow y$ removes all first derivative terms from the pressure operators.
Taking $d^2/dr^2 \to -k^2$, we have 
\be \label{eq:app-wkb}
 \omega & = & \underbrace{ \left(\frac{1+\gamma\beta^2}{1+\beta^2}\right) \left(-\frac{k^2c_{\rm iso}^2}{2\Omega}\right) + \omega_p}_{\text{pressure with $\beta$ cooling}}
 + \underbrace{\frac{\pi G \Sigma}{\Omega}|k| + \omega_g}_{\text{disk self-gravity}},
\ee
where
\be
 \omega_p & = & \frac{c_{\rm iso}^2}{2\Omega} \left[ \frac{1}{rP}\frac{d P}{dr} - \frac{1}{1+\beta^2}\left(\beta^2\gamma (r^3P)^{-1/2} \frac{d^2}{dr^2}(r^3P)^{1/2} + (r^3Pc_{\rm iso}^2)^{-1/2} \frac{d^2}{dr^2}(r^3Pc_{\rm iso}^2)^{1/2} \right)\right] \\
 \omega_g &=& -\frac{1}{2\Omega r^2}\frac{d}{dr}\left(r^2\frac{d\Phi_0}{dr}\right) - \left(\frac{\pi G \Sigma}{\Omega r}\right)\frac{3}{2\sqrt{k^2r^2 + 9/4}}  \left[ \frac{d}{d\ln{r}}\ln{\frac{\Sigma}{(r^3P)^{1/2}}}-\frac{1}{4}\right]
\ee
are the ``potential" profiles with corrections of order $|kr|^{-1}$. We have dropped all the imaginary terms in the final expressions. Our pressure part is different from \cite{Lee2019a,Lee2019b} because of cooling, but the self-gravity part is the same. 

\begin{figure}[htb!]
    \epsscale{0.6}
    \plotone{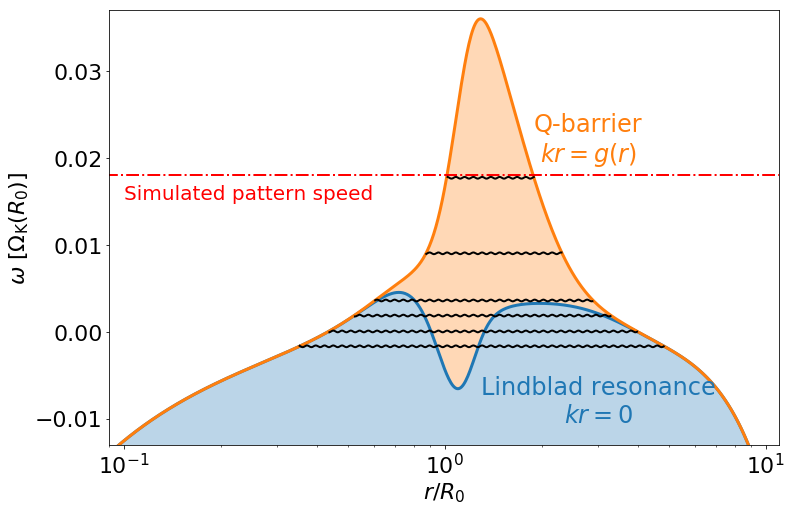}
    \caption{Frequency level diagram of the \textsc{Fiducial} disk. The black curves represent the first few standing modes from our WKB theory (fundamental mode at the top). Overplotted are the Lindblad resonances ($\omega|_{k=0}$; blue curve) and Q-barriers ($\omega|_{kr=g}$; orange curve).
    Intersection points between these curves and a given mode's $\omega$ mark the radial locations of that mode's Lindblad resonances or Q-barriers.Modes can only propagate in the shaded regions and are exponentially decaying outside these regions.
    The pattern speed from the simulation is shown in red to compare with the fundamental frequency.
    }
    \label{fig:app-WKB-QC}
\end{figure}

The DRMs (e.g., Figure~\ref{fig:Linear-DRM}) show the contours of $\omega$ in $r$-$kr$ space. Each contour that satisfies the quantum condition for standing waves as in \cite{Lee2019a} represents an eccentric mode allowed in the WKB theory.
Analogous to the energy level diagram of the solutions to wave functions in a potential well, Figure~\ref{fig:app-WKB-QC} is a frequency level diagram of the first few modes from the DRM of our \textsc{Fiducial} disk. The two modes at the top are the WKB ground state s-mode and its first harmonic. The ground state frequency shows excellent agreement with the simulated pattern speed. The rest are e-p/pg modes, which are characterized by their encounters with the Lindblad resonances.

\section{Growth rate}
\label{app:growth-rate}
The equation governing the modes angular momentum deficit can be obtained, to lowest order in $\beta$ and $d\beta/dr$, by multiplying Equation~\eqref{eq:Linear-Ecool} by $E^*$ and integrating over the disk,  
\begin{eqnarray}
    \label{eq:app-growth-rate-AMD}
    \int 2rL \omega |E|^2 dr &=& \int \bigg( \underbrace{ 
    \frac{E^*}{1 + \beta^2} \frac{d}{dr} \left\{ r^3 P \left[ c_{\rm iso}^2 \frac{d}{dr} \left(\frac{E}{c_{\rm iso}^2}\right) + \gamma \beta^2 \frac{dE }{dr}\right] + i \beta r^3 P \left[ (\gamma-1) \frac{dE}{dr} + \frac{d \ln c_{\rm iso}^2}{dr} E \right] \right\}
    + r^2 \frac{dP}{dr} |E|^2
    }_{\text{pressure with $\beta$ cooling}} 
    \nonumber \\ 
    &-&
    \overbrace{
    \Sigma r\frac{d}{dr}\big(r^2\frac{d\Phi_0}{dr}\big)|E|^2 - E^*\Sigma \frac{d}{dr}\big(r^2\Phi_1\big)
    }^{\text{disk self-gravity}}  + 
    \overbrace{i \frac{d\beta}{dr} r^2 P \left[ (\gamma-1) r E^*\frac{dE}{dr} + \frac{d\ln c_{\rm iso}^2}{d\ln r} |E|^2 \right]
    }^{\text{non-constant $\beta$}}  \bigg)dr,
\end{eqnarray}
where $L = r^2\OK\Sigma$. 

We first ignore the `non-constant $\beta$' contribution to derive the growth rate in Section~\ref{sec:Linear}. In the adiabatic limit, it has been shown that our form of self-gravity only contributes to the real part of $\omega$, i.e. the mode pattern speed \citep{Lee2019a}. Hence, the imaginary component of this equation is
\begin{eqnarray}
    \gamma_1 \int 2rL |E|^2 dr &=& \int \Im\bigg(
    \frac{E^*}{1 + \beta^2} \frac{d}{dr} \left\{ r^3 P \left[ c_{\rm iso}^2 \frac{d}{dr} \left(\frac{E}{c_{\rm iso}^2}\right) + \gamma \beta^2 \frac{dE }{dr}\right] + i \beta r^3 P \left[ (\gamma-1) \frac{dE}{dr} + \frac{d \ln c_{\rm iso}^2}{dr} E \right] \right\}\bigg)dr.
\end{eqnarray}
The right-hand side can be further simplified through integration by parts and assuming that the boundary contribution is zero:
\begin{eqnarray}
    \label{eq:app-growth-rate-AMD-imag}
    \gamma_1 \int 2rL |E|^2 dr &=& -\frac{1}{1 + \beta^2}\int \Im\bigg(
    \frac{dE^*}{dr}  \left\{ r^3 P c_{\rm iso}^2 \frac{d}{dr} \left(\frac{E}{c_{\rm iso}^2}\right)  + i \beta r^3 P \left[ (\gamma-1) \frac{dE}{dr} + \frac{d \ln c_{\rm iso}^2}{dr} E \right] \right\}\bigg)dr \nonumber \\
    &=& -\frac{1}{1 + \beta^2}\int r^3 P \left[\beta  (\gamma-1) |\frac{dE}{dr}|^2 - \Im((1-i\beta)\frac{d \ln c_{\rm iso}^2}{dr}E\frac{dE^*}{dr})\right]dr,
\end{eqnarray}
as Equation~\eqref{eq:Linear-AMDintegral} shows.

The correction due to the `non-constant $\beta$' part in Eqution~\eqref{eq:app-growth-rate-AMD} is
\begin{eqnarray}
    \Im(\omega_{\beta}) \int 2rL|E|^2 dr
    &=& \int r^2P \frac{d \beta}{d r}\left((\gamma-1)r\Re{(E^*\frac{dE}{dr})}+\frac{d\ln c_{\rm iso}^2}{d\ln r}|E|^2\right)dr.
\end{eqnarray}
Hence, with $dE/dr=ikE$ and $k = \gmax/r$, the growth rate is
\begin{eqnarray}
    \gamma_1 \approx -\frac{\gmax h^2 \Omega}{(2 \gamma)}[(\gamma-1) \beta\gmax + \frac{d \ln c_{\rm iso}^2}{d\ln r} {\rm sign}(k) - \frac{r}{\gmax}\frac{d\ln c_{\rm iso}^2}{d\ln r}\frac{d\beta}{dr}],
\end{eqnarray}
which has the criterion for instability:
\begin{eqnarray}
    \beta < \beta_{\rm c} = \frac{1}{g(\gamma-1)} \left(-\frac{d \ln c_{\rm iso}^2}{d \ln r}\right) \left[\text{sign}(k)-\frac{r}{g}\frac{d\beta}{dr}\right]
\end{eqnarray}
as mentioned in Section~\ref{sec:dis-beta}.

\section{Mode angular momentum deposition}
\label{app:tdep}

The torque from the mean flow to the wave is \citep{Dempsey2020} 
\begin{eqnarray}
    \frac{t_{\rm dep}}{2\pi r} = -\avg{u' \Sigma'}\pdr{r \avg{v}} +\avg{\Sigma} \avg{u' \pdr{r v'}}  - \avg{\frac{\Sigma'}{\Sigma} \pdp{P}'}.
\end{eqnarray}
The first two terms are related to the disk vortensity, while the third term is $\beta$ dependent and vanishes in the $\beta \rightarrow 0$ and $\beta \rightarrow+\infty$ limits.  

To evaluate the torque in term of the disk eccentricity, we first rewrite the first two terms as
\begin{eqnarray}
    \frac{t_{\rm dep,\xi}}{2\pi r} \equiv -\avg{u' \Sigma'}\pdr{r \avg{v}} +\avg{\Sigma} \avg{u' \pdr{r v'}} = 2\Sigma \Re(u'^*\pdr{(r v')}-u'^*\Sigma'r\xi),
\end{eqnarray}
where $\xi=\hat{\vec{z}} \cdot \vec{\nabla}\times\vec{v}/\Sigma = \partial_r (r v) / (r\Sigma)$ is the vortensity to the zeroth order. 

The disk vortensity evolution is governed by
\begin{eqnarray}
    \pdt{\xi} + u\pdr{\xi} + \frac{v}{r}\pdp{\xi} = \frac{1}{\Sigma^3}\vec{\nabla}\Sigma\times\vec{\nabla}P.
\end{eqnarray}
For a mode $\xi'\propto e^{i(\phi-\omega t)}$ with $|\omega|\ll\Omega$ and $\Im(\omega)=\gamma_1$, the governing equation can be simplified to
\begin{eqnarray}
    (i\Omega + \gamma_1) r\Sigma^2\xi' + u' r\Sigma^2 \pdr{\xi} = iTf' ,
    \label{eq:app-tdep-xievolution}\\
    {\rm where} \quad r\Sigma\xi' = \pdr{\ell'} - iu' - \Sigma'r\xi \quad {\rm and} \quad
    f' = \mu \Sigma \frac{T'}{T} - \eta\Sigma' ,
\end{eqnarray}
and where $\eta=d\ln T/dr$ and $\mu=d\ln \Sigma/dr$ for $T \equiv c^2_{\rm iso}$. Multiplying Equation~\eqref{eq:app-tdep-xievolution} by $(-i\Omega+\gamma_1)u'^*$, we get
\begin{eqnarray}
    (\Omega^2 + \gamma_1^2) r\Sigma^2 u'^*\xi' + (-i\Omega+\gamma_1)|u'|^2 r\Sigma^2 \pdr{\xi} = (\Omega+i\gamma_1)Tu'^*f'.
\end{eqnarray}
The real part of the above equation says
\begin{eqnarray}
\label{eq:app-tdepxi-Xprime}
    (\Omega^2 + \gamma_1^2)\frac{t_{\rm dep,\xi}}{4\pi r} + \gamma_1 r \Sigma^2 \pdr{\xi}|u'|^2 = \Omega T \Re(u'^*f') - \gamma_1 T \Im(u'^*f').
\end{eqnarray}
Hence, a non-zero $t_{\rm dep,\xi}$ can be caused by the growth/suppression of an eccentric mode (second LHS term) and the baroclinic effect (the RHS terms).

To proceed, we replace all primed quantities as 
\begin{eqnarray}
    u'=i\Omega r E,
    \quad
    \Sigma'=-(\frac{\Omega^2 + i\Omega\gamma_1}{\Omega^2 + \gamma_1^2})r\pdr{}(\Sigma E),
    \quad \text{and} \quad
    \frac{T'}{T} = i\beta(\gamma-1)\frac{\Sigma'}{\Sigma} - \frac{\beta}{\Omega}\pdr{S}u'.
\end{eqnarray}
The velocity $u'$ is from the eccentric formalism. $\Sigma'$ here is different from Equation~\eqref{eq:app-E-equation-uvS-E} because we need to keep the correction of order $\gamma_1$. The expression for $T'$ is from Equation~\eqref{eq:app-E-equation-SE} in the limit of $\beta\ll1$. Plugging them into Equation~\eqref{eq:app-tdepxi-Xprime} gives
\begin{eqnarray}
    (\Omega^2 + \gamma_1^2)\frac{t_{\rm dep,\xi}}{4\pi r} = - \gamma_1\Omega^2r^3\Sigma^2\pdr{\xi}|E|^2 + \Omega^2 r^2 \Sigma \pdr{T} \left[ \Im(E^*\pdr{E}) 
    -\frac{\beta \mu^2}{\eta}\frac{\gamma-1}{2}\pdr{|E|^2}
    -\mu\beta|E|^2
    \right],
\end{eqnarray}
where we drop the small terms proportional to $\gamma_1\beta$ and $\gamma_1 T$. 
 
The contribution from the cooling term,
\begin{eqnarray}
    \frac{t_{\rm dep,\beta}}{2\pi r} \equiv - \avg{\frac{\Sigma'}{\Sigma} \pdp{P}'} = -2\Re(i\Sigma'^*T'),
\end{eqnarray}
can be rewritten as
\begin{eqnarray}
    \frac{t_{\rm dep,\beta}}{4\pi r} = \beta r^2 \Sigma T \left[ \mu \eta |E|^2 
    + (\gamma-1)\mu\pdr{|E|^2}
    + (\gamma-1)\left|\pdr{E}\right|^2\right]
\end{eqnarray}
with the same substitutions that we used for $t_{\rm dep,\xi}$.

Therefore, the angular momentum transferred from the mean flow to the wave is 
\begin{eqnarray}
    \frac{t_{\rm dep}}{4\pi r} &=& \frac{t_{\rm dep,\xi}}{4\pi r} + \frac{t_{\rm dep,\beta}}{4\pi r} \\
    &=& 
    - \gamma_1 r^3 \Sigma^2 \pdr{\avg{\xi}} |E|^2 
    + r^2 \Sigma \pdr{T}  \left[\Im\left(E^* \pdr{E}\right)-\frac{\beta \mu^2}{\eta}\frac{\gamma-1}{2}\pdr{|E|^2}\right] 
    + \beta  (\gamma-1) r^2 \Sigma T \left[\mu\pdr{|E|^2}+ \left|\pdr{E}\right|^2\right].
\end{eqnarray}

For Equation~\eqref{eq:m0-tdep-E} in the main text, we drop terms proportional to $\partial_r|E|^2$ in the brackets above. This is because when $E\propto e^{ikr}$ and $|kr| \gg 1$,
\be
\partial_r|E|^2 \sim \Re(2ik|E|^2) \, \lesssim \,
\Im(E^*\partial_r E) \sim k|E|^2 \, \lesssim \, 
|\partial_r E|^2\sim k^2 |E|^2,
\ee
so that the $\partial_r|E|^2$ term has the smallest order of magnitude.

%\bibliography{cool-disk}{}
%\bibliographystyle{aasjournalnolink}

\end{document}